\documentclass[11pt,a4paper]{article}
\usepackage[a4paper, twoside, left=21mm, right=21mm, top=24mm, bottom=24mm]{geometry}
\usepackage{amsmath}
\usepackage{braket}
\usepackage{authblk}
\usepackage{mathrsfs}
\usepackage{bbm}
\usepackage{graphicx}
\usepackage{subcaption}
\usepackage{slashed}
\usepackage{cite}
\usepackage{booktabs}
\usepackage[colorlinks=true, citecolor=green, linkcolor=blue, urlcolor=red]{hyperref}
\title{Spin correlation of hyperon-antihyperon systems in $e^+e^-$ annihilation}
\author[1]{Wenyu~Zhang\thanks{zwy\_2020@mail.ustc.edu.cn}}
\author[1, 2, *]{Yang~Li\thanks{leeyoung1987@ustc.edu.cn}}
\author[1,3]{Xiaorong~Zhou\thanks{zxrong@ustc.edu.cn}}
\author[1,2,4]{Qun~Wang\thanks{qunwang@ustc.edu.cn}}
\affil[1]{\textit{Department of Modern Physics, University of Science and Technology of China, Hefei 230026, China}}
\affil[2]{\textit{Anhui Center for fundamental sciences in theoretical physics, Hefei 230026, China}}

\affil[3]{\textit{State Key Laboratory of Particle Detection and Electronics, Beijing 100049, Hefei 230026, China}}
\affil[4]{\textit{School of Mechanics and Physics, Anhui University of Science and Technology, Huainan, Anhui 232001, China}}

\begin{document}
\maketitle

\begin{abstract}
Quantum correlations in high-energy collisions provide a novel perspective on both fundamental physics and hadron structure. We calculate the quantum information observable in spin-1/2 particle-antiparticle systems produced in $e^+e^-$ annihilation. Starting from the two-qubit density operator, we calculate the Bell variable, concurrence and negativity as functions of the scattering angle and collision energy for the $\Lambda\bar{\Lambda}$, $\Sigma^+\bar{\Sigma}^-$, and $\Lambda^+_c\bar{\Lambda}^-_c$ systems in the process $e^+e^-\rightarrow B\bar{B}$ at BESIII experiments. Unlike elementary particle-antiparticle systems, the hyperon-antihyperon system exhibits non-vanishing transverse polarization for $B$ and $\bar{B}$ with respect to the production plane, which significantly restrict the kinematic region where the CHSH inequality is violated while leaving the entanglement largely unaffected. We also extend our analysis to the general case of spin-1/2 particle-antiparticle production, and explore the potential of using quantum correlations as probes to hadron structure, particularly to the parton-level entanglement inside heavy-flavored mesons. 

\end{abstract}

\section{Introduction}

Quantum information theory provides a powerful framework for characterizing nonclassical correlations, including Bell nonlocality and quantum entanglement. Demonstrated through the violation of Bell-type inequalities~\cite{Bell:1964kc,Clauser:1974tg,Aspect:1981zz}, Bell nonlocality reveals correlations that defy explanation by any local hidden-variable theory~\cite{Brunner:2013est}. Meanwhile, quantum entanglement describes non-separable quantum states in which the properties of individual particles remain intrinsically linked, irrespective of their spatial separation~\cite{Horodecki:2009zz}. This phenomenon underpins practical applications across quantum information processing, ranging from quantum computing~\cite{Jozsa:2002rcj}, metrology~\cite{Giovannetti:2011chh}, and communication~\cite{Curty:2003ybi} to quantum machine learning~\cite{Hentschel:2010rbn}. Together with concepts like quantum steering~\cite{Wiseman:2007hyt,Uola:2020kps} and discord~\cite{Ollivier:2001fdq}, these phenomena form a hierarchy that quantifies the degree of ``quantumness'' inherent in a physical system. Although traditionally investigated in highly controlled photonic and atomic systems~\cite{Yin:2017iwp,BIGBellTest:2018ebd}, the study of these phenomena has recently expanded into high-energy physics within the framework of quantum field theory. These extreme energies and clean experimental environments offer unique opportunities to generate and probe entangled systems of elementary particles, forging a new interface between high-energy physics and quantum information science.

Quantum correlations in high-energy particle and nuclear physics have been extensively investigated in recent years. 
Top-quark pairs have proven particularly well-suited for these investigations; owing to the top quark's exceptionally short lifetime, its spin information is transmitted directly to its decay products before hadronization can occur~\cite{Afik:2020onf,Aguilar-Saavedra:2023hss,Cheng:2023qmz,Aguilar-Saavedra:2024hwd,Cheng:2024btk}. Recently, both the ATLAS and CMS collaborations reported the successful observation of entanglement in top-quark events at the Large Hadron Collider (LHC), representing a major milestone in applying quantum information concepts to high-energy physics~\cite{ATLAS:2023fsd,CMS:2024pts}. Beyond top quarks, theoretical proposals and experimental studies have been extended to other systems, including tau-lepton pairs~\cite{Fabbrichesi:2022ovb,Zhang:2025mmm}, bottom-quark pairs~\cite{Afik:2025grr}, and massive gauge bosons originating from Higgs boson decays~\cite{Aguilar-Saavedra:2022wam,Aguilar-Saavedra:2022mpg,Aguilar-Saavedra:2024whi,Fabbrichesi:2023cev,Aoude:2023hxv,Bernal:2023ruk,DelGratta:2025qyp,Goncalves:2025xer}. Furthermore, lepton colliders provide excellent environments for probing spin correlations in final-state systems such as $t\bar{t}$~\cite{Maltoni:2024csn}, $\tau^+\tau^-$~\cite{Altakach:2022ywa,Ma:2023yvd,Ehataht:2023zzt,LoChiatto:2024dmx,Han:2025ewp}, $\mu^+\mu^-$~\cite{Ruzi:2024iqu}, $\phi\phi$~\cite{Gabrielli:2024kbz}, and dibosons~\cite{Ruzi:2024cbt,Wu:2024ovc,Ding:2025mzj}. Collectively, these investigations demonstrate that various fermion and boson pairs produced in high-energy collisions serve as accessible and powerful laboratories for examining fundamental quantum phenomena.

The STAR Collaboration recently reported the first experimental evidence of $\Lambda\bar{\Lambda}$ spin correlations in high-energy proton–proton collisions at the Relativistic Heavy-Ion Collider (RHIC) \cite{STAR:2025njp}. These correlations are inherited from quark–antiquark pairs produced from the QCD vacuum. The spin correlation and quantum entanglement of this $\Lambda\bar{\Lambda}$ system have been theoretically studied via string fragmentation models, where successive $q\bar{q}$ pairs of various flavors emerge in spin-singlet states \cite{Gong:2021bcp}. Furthermore, in heavy-ion collisions, STAR observed a significant spin alignment of $\phi$ mesons \cite{STAR:2022fan}, implying a spin correlation between the constituent strange quark and antiquark \cite{Sheng:2019kmk,Sheng:2022ffb,Chen:2024afy}. Such constituent-quark spin correlations have been incorporated into quark coalescence models \cite{Lv:2024uev}, offering a potential explanation for the possible excess in the global polarization of $\Omega$ hyperons relative to other hyperon species \cite{Oliva:2026wbo}.

Following its recent upgrade, the BESIII experiment at the Beijing Electron-Positron Collider (BEPC), which operates at energies near the charmonium mass region, offers a pristine environment for studying quantum correlations in hyperon-antihyperon systems~\cite{Wu:2024mtj,Wu:2024asu,Wu:2025dds,Fabbrichesi:2024rec,BESIII:2025vsr,Tang:2025oav,Jaloum:2025mhh,Liu:2026tsw,Zhang:2026nwm,Li:2026bkf,Jaloum:2026bnp}. Unlike elementary leptons, hyperons are composite particles whose weak decays serve as highly effective spin polarimeters, enabling the complete reconstruction of their spin density matrices. This inherent feature has facilitated detailed investigations of entanglement and Bell nonlocality in hyperon-antihyperon pairs produced via charmonium decays. Furthermore, the production of these pairs in electron-positron annihilation involves time-like electromagnetic form factors that can induce a net polarization in the final state~\cite{Faldt:2016qee,Faldt:2017kgy}, a characteristic completely absent in systems such as $\tau^+\tau^-$ or $t\bar{t}$. This distinct polarization effect sets the hyperon-antihyperon system apart, providing a unique setting to explore how single-spin observables influence various forms of quantum correlations.

In this paper, we present a systematic study of quantum correlations in spin-1/2 particle-antiparticle systems produced via $e^+e^-$ annihilation, focusing on hyperon pairs ($\Lambda\bar{\Lambda}$, $\Sigma^+\bar{\Sigma}^-$, and $\Lambda^+_c\bar{\Lambda}^-_c$) accessible at the BESIII experiment~\cite{BESIII:2023ynq,BESIII:2025kyg,BESIII:2025zbz}. We perform a quantum tomographic analysis of the hyperon-antihyperon system by employing the two-qubit density operator~\cite{Perotti:2018wxm,Batozskaya:2023rek}. Through this framework, we calculate three quantum information observables, the Bell variable $\mathcal{B}$ for nonlocality, alongside the concurrence $\mathcal{C}$ and negativity $\mathcal{N}$ for entanglement, as functions of the scattering angle $\theta$. For the first time, we also investigate their dependence on the center-of-mass energy $\sqrt{s}$. We then extend our analysis to the general case of spin-1/2 particle-antiparticle systems to explore the hierarchical relationships among these quantum resources, detailing how final-state polarization affects nonlocality differently than it does entanglement. Finally, we discuss the implications of using these quantum correlations as sensitive probes of hadron structure, highlighting the potential to experimentally access parton-level entanglement within heavy-flavored mesons.

The paper is organized as follows. Section~\ref{Sec:qtqiob} starts from the two-qubit density operator and introduces the quantum information observables. Section~\ref{Sec:results} presents the numerical results for these observables in hyperon-antihyperon systems in $e^+e^-$ annihilation as wells in general $B\bar{B}$ systems. The connections of these observables with the parton entanglement in heavy-flavored mesons are discussed. Section~\ref{Sec:Summary and outlook} presents a summary of the main results and an outlook for future work.

\section{Quantum information observables}
\label{Sec:qtqiob}
In this section, we present the general form of the spin density operator for spin-1/2 particle-antiparticle systems and introduce the quantum information observables including the concurrence, negativity, and Bell variable. 

\subsection{Spin density matrix} 
The baryon-antibaryon $B\bar{B}$ system in the process $e^+e^-\rightarrow \gamma^*\rightarrow B\bar{B}$ forms a two-qubit system. Their spin states can be described by the spin density operator $\rho_{B\bar{B}}$ in the Fano-Bloch decomposition \cite{Fano:1983zz}  
\begin{equation}
    \rho_{B\bar{B}}=\frac{1}{4}\bigg(\mathbbmtt{1}\otimes\mathbbmtt{1}+\boldsymbol{\sigma}\cdot\mathbf{P}^+
    \otimes\mathbbmtt{1}+\mathbbmtt{1}\otimes\boldsymbol{\sigma}\cdot\mathbf{P}^-
    +C_{ij}\sigma_{i}\otimes\sigma_{j}\bigg),
\label{eqa:spin density operator}
\end{equation}
where $\mathbbmtt{1}$ is the $2\times 2$ identity matrix, $\boldsymbol{\sigma}=(\sigma_1,\sigma_2,\sigma_3)=(\sigma_x,\sigma_y,\sigma_z)$ are Pauli matrices, $\mathbf{P}^+$ and $\mathbf{P}^-$ are the polarization vectors of $B$ and $\bar{B}$ respectively, $C_{ij}$ is the $3\times3$ spin correlation matrix between them. 

In the process $\gamma^*\rightarrow B\bar{B}$, the electromagnetic current can be expressed in terms of the baryon's time-like form factors $F_1$ and $F_2$ as
\begin{equation}
\Gamma^{\mu}=\gamma^{\mu}F_1(q^2)+i\frac{\sigma^{\mu\nu}q_{\nu}}{2M}F_2(q^2),
\label{em-current}
\end{equation}
where $q=p_1+p_2$ with $p_1$ and $p_2$ being the four-momentum of $B$ and $\bar{B}$ respectively, and $M$ is the baryon mass. These form factors are related to the time-like electric and magnetic form factors $G_E$ and $G_M$ by
\begin{equation}
G_E=F_1+\frac{q^2}{4M^2}F_2,\qquad G_M=F_1+F_2.
\end{equation}
With the electromagnetic current (\ref{em-current}), the spin polarization vectors and spin correlation matrix in the rest frame of $B$ and $\bar{B}$ are given by~\cite{Perotti:2018wxm,Batozskaya:2023rek}
\begin{align}
    \mathbf{P}^+=&\mathbf{P}^-=\bigg(0,\frac{\beta\operatorname{sin}\theta\operatorname{cos}\theta}{1+\alpha\operatorname{cos}^2\theta},0\bigg), 
\label{eqa:polarization} \\
    C_{ij}=&\frac{1}{1+\alpha \operatorname{cos}^2\theta}
    \begin{pmatrix}
        \operatorname{sin}^2\theta & 0 & \gamma \operatorname{sin}\theta \operatorname{cos}\theta \\
        0 & -\alpha \operatorname{sin}^2\theta & 0 \\
        \gamma \operatorname{sin}\theta \operatorname{cos}\theta & 0 & \alpha+\operatorname{cos}^2\theta
    \end{pmatrix}. 
\label{eqa:spin-correlation-matrix}
\end{align}
Here, $\theta$ is the scattering angle between the electron beam and the outgoing particle, and $\alpha$, $\beta$ and $\gamma$ are three parameters defined through $G_E$ and $G_M$ as  
\begin{align}
    \alpha=&\frac{s|G_M|^2-4M^2|G_E|^2}{s|G_M|^2+4M^2|G_E|^2}\in[-1,1],\nonumber\\
    \beta=&\sqrt{1-\alpha^2}\operatorname{sin}(\Delta\Phi),\nonumber\\
    \gamma=&\sqrt{1-\alpha^2}\operatorname{cos}(\Delta\Phi), \nonumber\\
    \Delta\Phi=&\mathrm{arg}\left(\frac{G_E}{G_M}\right)\in(-\pi,\pi],
\end{align}
where $s=q^2$ is the center-of-mass energy squared and $\Delta\Phi$ is the phase of $G_E/G_M$. 
The three basis vectors of the coordinate system are chosen to be 
\begin{equation}
    \hat{\textbf{y}}=\frac{\hat{\textbf{p}}_{B}\times\hat{\textbf{p}}_e}{|\hat{\textbf{p}}_{B}\times\hat{\textbf{p}}_e|},\qquad\hat{\textbf{z}}=\hat{\textbf{p}}_B,\qquad\hat{\textbf{x}}=\hat{\textbf{y}}\times\hat{\textbf{z}},
\end{equation}
which applies to both $B$ and $\bar{B}$. 

\subsection{Entanglement observables}
The quantum state of a composite system $AB$ can be described by a density operator $\rho_{AB}$ acting on the tensor product Hilbert space $\mathcal{H}_A \otimes \mathcal{H}_B$. A pure state $\ket{\Psi}_{AB}$ is said to be separable (i.e., unentangled) if it can be written as a product state $\ket{\psi}_A \otimes \ket{\phi}_B$. Otherwise, it is entangled. For mixed states, separability is defined more complicated: $\rho_{AB}$ is separable if it can be expressed as a convex combination of product states,
\begin{equation}
\rho_{AB} = \sum_k p_k \, \rho_A^{(k)} \otimes \rho_B^{(k)}, \quad p_k \geq 0, \quad \sum_k p_k = 1.
\end{equation}
Any state that cannot be written in this form is entangled. Detecting entanglement in practice is nontrivial, especially when the full density matrix is unknown. An entanglement witness is an observable $\mathcal{W}$ such that $\operatorname{Tr}(\mathcal{W} \rho_{\text{sep}}) \geq 0$ for all separable states $\rho_{\text{sep}}$, but $\operatorname{Tr}(\mathcal{W} \rho_{\text{ent}}) < 0$ for at least one entangled state $\rho_{\text{ent}}$, see, for example, Refs.~\cite{Horodecki:2009zz, Guhne:2008qic} for reviews on this topics with more details.

For a pure bipartite state $\ket{\Psi}_{AB}$, the canonical measure of entanglement is the {entanglement entropy}, defined as the von Neumann entropy $S_\text{vN}$ of the reduced density matrix of either subsystem:
\begin{equation}
S_A \equiv S_\text{vN}(\rho_A) = -\operatorname{Tr}_A \left( \rho_A \log \rho_A \right), \quad \text{where} \quad \rho_A = \operatorname{Tr}_B \left( \ket{\Psi}\bra{\Psi} \right).
\end{equation}
However, the spin density operator $\rho_{B\bar{B}}$ in this work is for a mixed state. To explicitly detect the quantum entanglement between spin states of $B$ and $\bar{B}$, the von Neumann entropy is no longer applicable, we use in this paper robust entanglement witnesses: the negativity from PPT criterion and concurrence.

For two-qubit and qubit-qutrit systems (corresponding to Hilbert space $2\otimes2$ and $2\otimes3$ respectively), the Peres-Horodecki criterion provides a sufficient and necessary condition for separability~\cite{Peres:1996dw,Lewenstein:2000jlo} of a bipartite state 
\begin{equation}
    \rho_{AB} = \sum_{ijkl} c_{ijkl}\ket{i}_A \ket{j}_B \bra{k}_A \bra{l}_B
\end{equation}
is separable if and only if its partial transpose
\begin{equation}
    \rho_{AB}^{T_B} = \sum_{ijkl} c_{ijkl} \ket{i}_A \ket{l}_B \bra{k}_A \bra{j}_B
\end{equation}
for the second subsystem is positive semi-definite. The Peres-Horodecki criterion is also called the Positive Partial Transpose (PPT) criterion. Closely related to the PPT criterion is the quantum negativity~\cite{Vidal:2002zz}, defined as
\begin{equation}
    \mathcal{N}[\rho_{AB}]=\frac{1}{2}\Big(||\rho^{T_B}_{AB}||_1-1\Big),
\end{equation}
where $||\cdot||_1$ denotes the trace norm. The negativity quantifies entanglement by summing the absolute values of the negative eigenvalues of the partially transposed matrix. A larger negativity indicates a greater violation of the PPT criterion. Especially, for a two-qubit system, negativity is in the range $[0,0.5]$. A state is separable if $\mathcal{N}[\rho]=0$ and is entangled if $\mathcal{N}[\rho]>0$. When $\mathcal{N}[\rho]=0.5$, the state is said to be maximally entangled. 

The concurrence is another entanglement monotone, which has a direct relationship with entanglement of formation~\cite{Hill:1997pfa}. In Ref.~\cite{Wootters:1997id}, Wootters derived the two-qubit concurrence as
\begin{equation}
    \mathcal{C}[\rho]=\text{max}\{0,\mu_1-\mu_2-\mu_3-\mu_4\},
\end{equation}
where $\mu_i$ with $i=1,2,3,4$ are the eigenvalues of the Hermitain matrix $\sqrt{\sqrt{\rho}\tilde{\rho}\sqrt{\rho}}$ in the decreasing order, with $\tilde{\rho}$ being defined as 
\begin{equation}
    \tilde{\rho}=(\sigma_y\otimes\sigma_y)\rho^*(\sigma_y\otimes\sigma_y)\;,
\end{equation}
and $\rho^*$ being the complex conjugate of $\rho$. Wooters' concurrence is a function in the range $[0,1]$. Similar to the negativity, a state is sparable if $\mathcal{C}[\rho]=0$ and it is entangled if $\mathcal{C}[\rho]>0$. When $\mathcal{C}[\rho]=1$, the state is said to be maximally entangled.

\subsection{Bell variable}
\label{sec:Bell-nonlocality}
Another way to detect the entanglement is through Bell inequality. For a two-qubit system, the most widely used Bell-type inequality is the CHSH inequality~\cite{Clauser:1969ny}. In the particle-antiparticle spin correlation system, it can be written as
\begin{equation}
    \left|\langle \boldsymbol{a}_1\cdot \boldsymbol{\sigma}\otimes \boldsymbol{b}_1\cdot \boldsymbol{\sigma}\rangle+
    \langle \boldsymbol{a}_1\cdot \boldsymbol{\sigma}\otimes \boldsymbol{b}_2\cdot \boldsymbol{\sigma}\rangle+
    \langle \boldsymbol{a}_2\cdot \boldsymbol{\sigma}\otimes \boldsymbol{b}_1\cdot \boldsymbol{\sigma}\rangle-
    \langle \boldsymbol{a}_2\cdot \boldsymbol{\sigma}\otimes \boldsymbol{b}_2\cdot \boldsymbol{\sigma}\rangle\right| \leq 2, 
\label{chsh}
\end{equation}
where $\braket{\mathcal{O}}\equiv\operatorname{Tr}[\rho\mathcal{O}]$, 
$\boldsymbol{a}_1$ and $\boldsymbol{a}_2$ specify the "detector settings" or the directions along which the spin polarization of the particle $B$ is measured, and $\boldsymbol{b}_1$ and $\boldsymbol{b}_2$ specify the "detector settings" or the directions along which the spin polarization of antiparticle $\bar{B}$ is measured. The Bell variable $\mathcal{B}$ measures the maximum of the left-hand side of the CHSH inequality (\ref{chsh}) among all different choice of $\boldsymbol{a}_1$, $\boldsymbol{a}_2$, $\boldsymbol{b}_1$, $\boldsymbol{b}_2$. The optimal set of choices that maximize the left-hand side of the CHSH inequality (\ref{chsh}) yields a value 
\begin{equation}
    \mathcal{B}[\rho]=2\sqrt{m_1+m_2},
\label{eqa:Bell variable}
\end{equation}
where $m_1$ and $m_2$ are two largest eigenvalues of the matrix $C^TC$~\cite{Horodecki:1995nsk}. The CHSH inequality can be violated if and only if $\mathcal{B}[\rho]>2$, and the maximum possible violation of the CHSH inequality is the upper bound value $2\sqrt{2}$. 

It is worth noting that although the Bell inequality can be used to test the entanglement, the value of $\mathcal{B}[\rho]$ can not serve as an entanglement measure. It can only quantify the degree of nonlocality in a system~\cite{Werner:1989zz,Munro:2001gth}. On one hand, a two-qubit system that is entangled may still obey the Bell inequality, on the other hand, even if $\mathcal{B}[\rho]$ is significantly greater than 2 the system may still be weakly entangled. Therefore, the Bell inequality can only be used to determine the presence or absence of quantum entanglement, and cannot be employed as a measure of its magnitude. 

\section{Numerical results and discussions} 
\label{Sec:results}
In this section, we present the results for quantum information observables in hyperon-antihyperon systems in the process $e^+e^-\rightarrow \gamma^* \rightarrow B\bar{B}$ ($B=\Sigma^+$, $\Lambda$ and $\Lambda_c^+$), and explore how they depend on the center-of-mass energy $\sqrt{s}$. We then discuss different quantum information observables in hyperon-antihyperon systems and in general $B\bar{B}$ systems, as well as implications of these results for the experimental measurement of the parton entanglement in heavy-flavored mesons.

\subsection{Numerical results}
Our analysis relies on the parameters in the spin density operator provided by the BESIII collaboration. The analytical formulas for the quantum information observables are given in Appendix~\ref{appendix:formula-ob}.
Tables \ref{tab:Sigma data}-\ref{tab:Lambda_c data} list the values of $\alpha$ and $\Delta\Phi$ for the processes $e^+e^-\rightarrow B\bar{B}$ ($B=\Sigma^+$, $\Lambda$ and $\Lambda_c^+$) at different energies. 

\begin{table}[ht]
\centering
\caption{Parameters in $e^+e^-\rightarrow\Sigma^+\bar{\Sigma}^-$ at some $\sqrt{s}$~\cite{BESIII:2023ynq}.}
\begin{tabular}{cccc}
\toprule
\( \sqrt{s}/\text{MeV} \) & $\tau_B$ & \( \alpha \) & \( \Delta \Phi / \text{rad} \) \\
\hline
\( 2396 \) & \(0.015\) & $-0.48\pm0.18\pm0.09$ & $-0.733\ (-2.409)\pm0.384\pm0.244$ \\
\( 2645 \) & \(0.237\) & $0.41\pm0.12\pm0.06$ & $0.960\pm0.332\pm0.244$ \\
\( 2900 \) & \(0.486\) & $0.35\pm0.17\pm0.15$ & $1.361\pm0.384\pm0.157$ \\
\bottomrule
\label{tab:Sigma data}
\end{tabular}
\end{table}

\begin{table}[ht]
\centering
\caption{Parameters in $e^+e^-\rightarrow\Lambda\bar{\Lambda}$ at some $\sqrt{s}$~\cite{BESIII:2025kyg}.}
\begin{tabular}{cccc}
\toprule
\( \sqrt{s}/\text{MeV} \) & $\tau_B$ &  \( \alpha \) & \( \Delta \Phi / \text{rad} \)\\
\hline
\( 2386 \) & \(0.144\) & $-0.010\pm0.139\pm0.028$ & $0.646\pm0.209\pm0.017$ \\
\( 2396 \) & \(0.153\) & $0.164\pm0.086\pm0.021$ & $0.733\pm0.175\pm0.035$ \\
\( 2645 \) & \(0.406\) & $0.112\pm0.149\pm0.028$ & $2.566\pm0.227\pm0.017$ \\
\( 2900 \) & \(0.689\) & $0.219\pm0.229\pm0.027$ & $2.583\pm0.332\pm0.017$ \\
\( 3080 \) & \(0.905\) & $0.619\pm0.304\pm0.028$ & $1.396\pm1.204\pm0.035$ \\
\bottomrule
\label{tab:Lambda data}
\end{tabular}
\end{table}

\begin{table}[ht]
\centering
\caption{Parameters in $e^+e^-\rightarrow\Lambda_c^+\bar{\Lambda}_c^-$ at some $\sqrt{s}$~\cite{BESIII:2025zbz}.}
\begin{tabular}{cccc}
\toprule
\( \sqrt{s}/\text{MeV} \) & $\tau_B$ & \( \alpha \) & \( \Delta \Phi / \text{rad} \) \\
\hline
\( 4600 \) & \(0.012\) & $-0.226\pm0.030\pm0.004$ & $0.100\pm0.069\pm0.009$ \\
\( 4612 \) & \(0.017\) & $-0.160\pm0.083\pm0.004$ & $0.146\pm0.162\pm0.030$ \\
\( 4628 \) & \(0.024\) & $-0.181\pm0.038\pm0.001$ & $0.371\pm0.082\pm0.012$ \\
\( 4641 \) & \(0.030\) & $-0.060\pm0.039\pm0.003$ & $0.398\pm0.073\pm0.015$ \\
\( 4661 \) & \(0.039\) & $0.008\pm0.044\pm0.003$ & $0.496\pm0.088\pm0.021$ \\
\( 4682 \) & \(0.048\) & $0.102\pm0.029\pm0.003$ & $0.502\pm0.054\pm0.021$ \\
\( 4699 \) & \(0.056\) & $0.305\pm0.055\pm0.010$ & $0.545\pm0.114\pm0.028$ \\
\( 4740 \) & \(0.074\) & $0.358\pm0.126\pm0.008$ & $0.097\pm0.190\pm0.016$ \\
\( 4750 \) & \(0.079\) & $0.347\pm0.079\pm0.004$ & $0.316\pm0.142\pm0.019$ \\
\( 4781 \) & \(0.093\) & $0.157\pm0.062\pm0.007$ & $0.395\pm0.126\pm0.028$ \\
\( 4843 \) & \(0.122\) & $0.282\pm0.089\pm0.019$ & $0.385\pm0.153\pm0.034$ \\
\( 4918 \) & \(0.157\) & $0.612\pm0.150\pm0.019$ & $0.423\pm0.272\pm0.024$ \\
\( 4951 \) & \(0.172\) & $0.744\pm0.179\pm0.007$ & $0.700\pm0.392\pm0.058$ \\
\bottomrule
\label{tab:Lambda_c data}
\end{tabular}
\end{table}

\begin{figure}[p]
    \centering
    \begin{subfigure}{0.32\textwidth}
        \includegraphics[width=\textwidth]{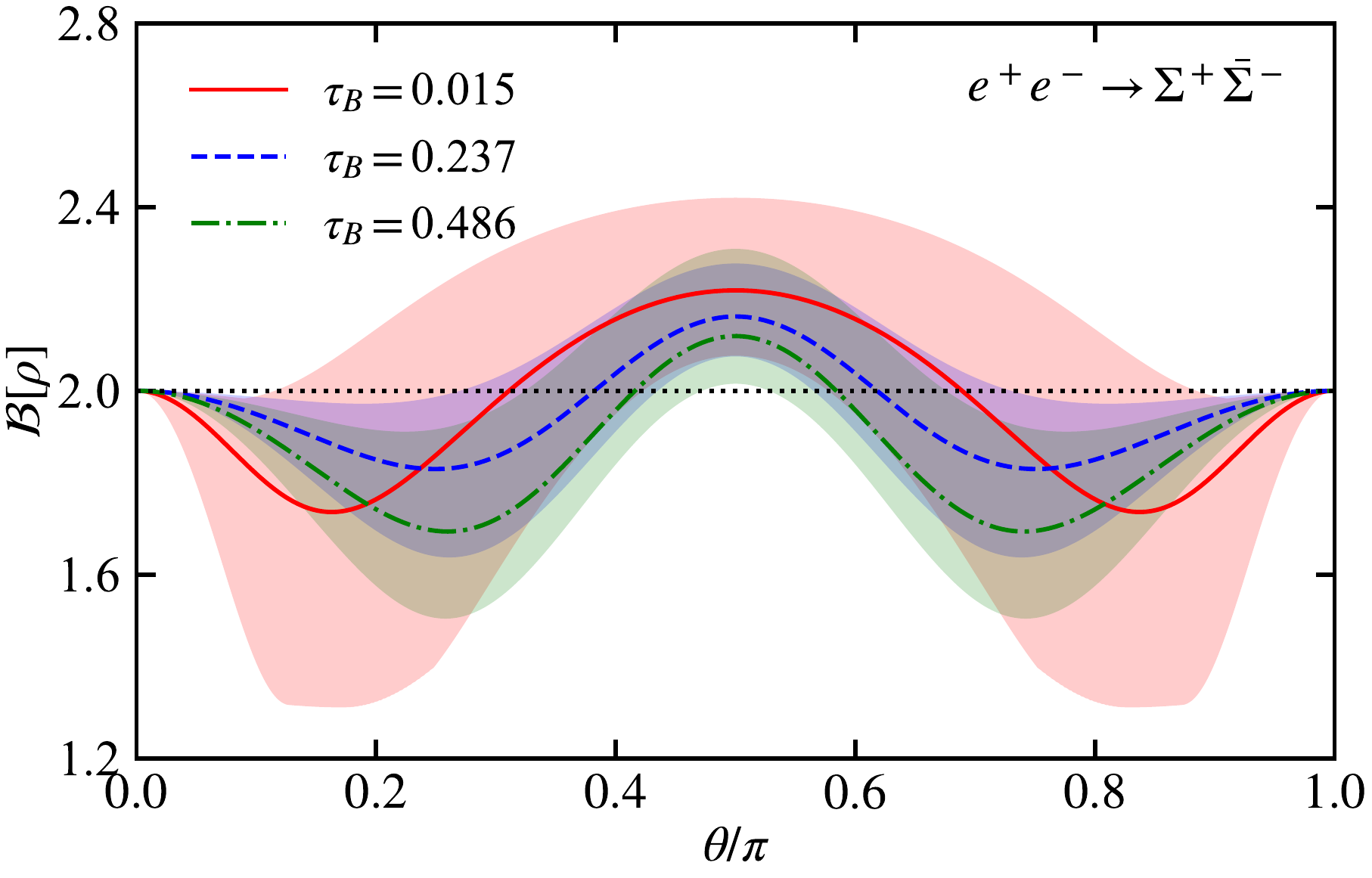}
        \caption{}
    \label{BESIII:a}
    \end{subfigure}
    \hfill
    \begin{subfigure}{0.32\textwidth}
        \includegraphics[width=\textwidth]{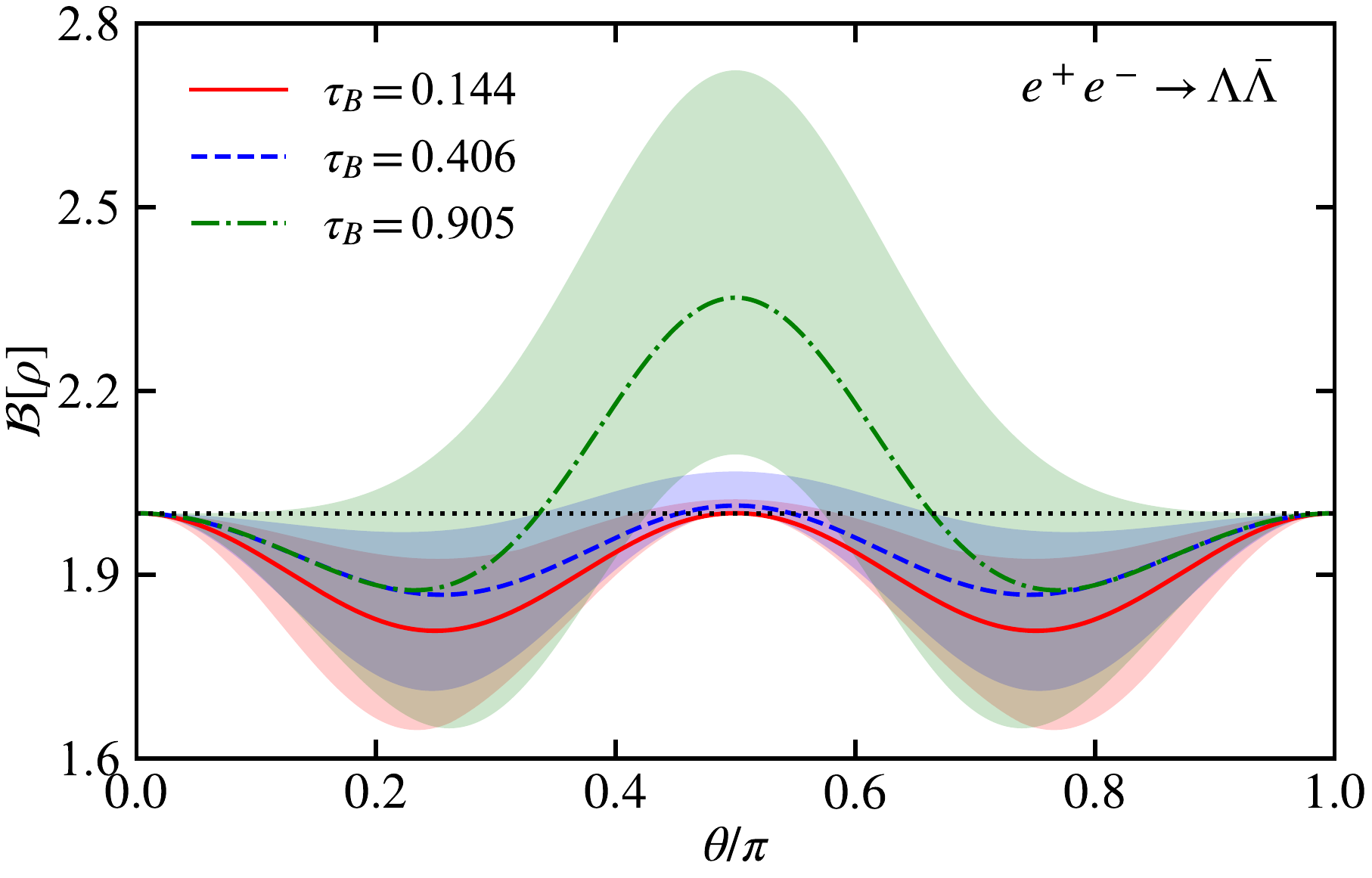}
        \caption{}
    \label{BESIII:b}
    \end{subfigure}
    \hfill
    \begin{subfigure}{0.32\textwidth}
        \includegraphics[width=\textwidth]{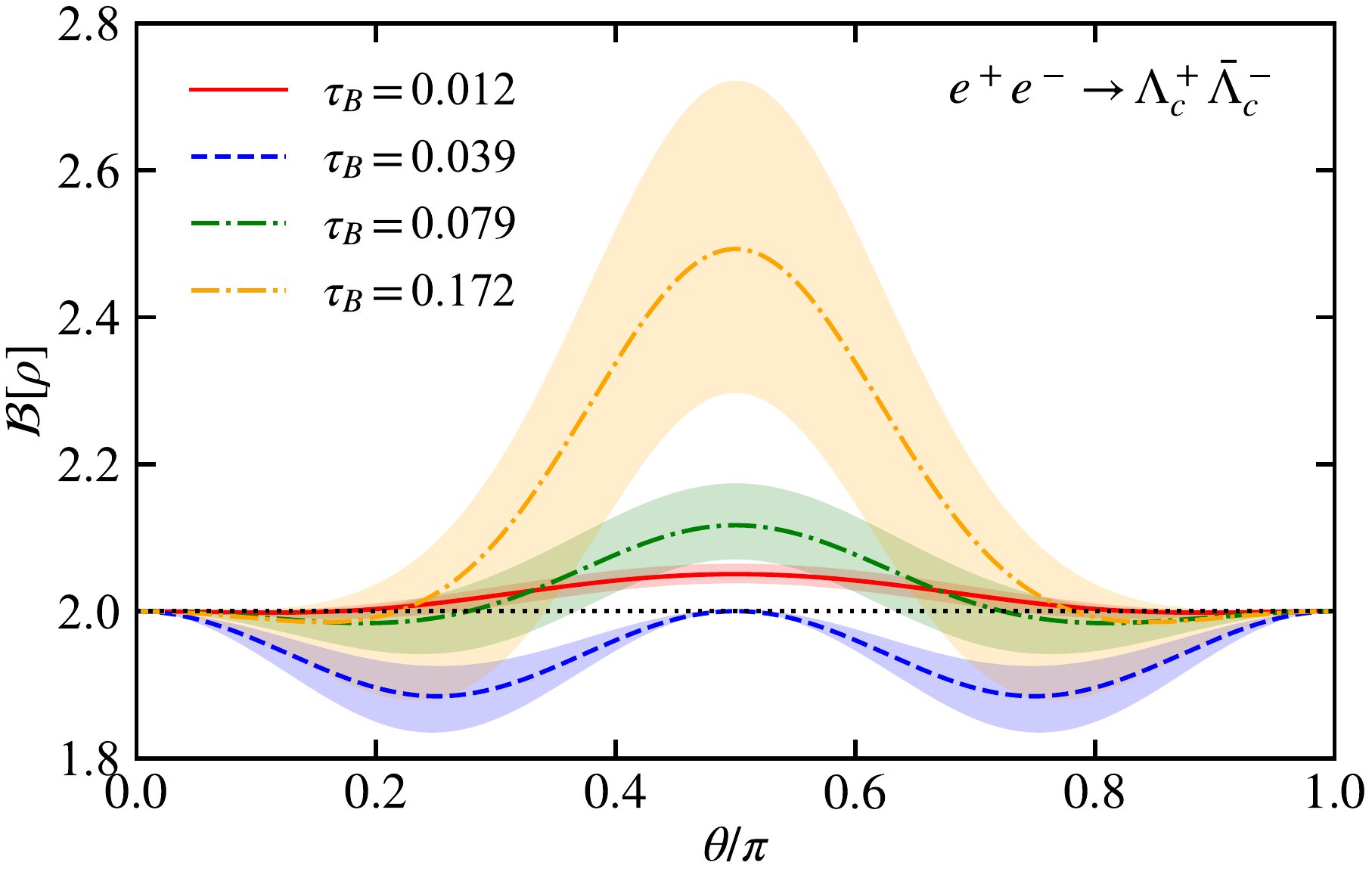}
        \caption{}
    \label{BESIII:c}
    \end{subfigure}
    \vskip\baselineskip
    \centering
    \begin{subfigure}{0.32\textwidth}
        \includegraphics[width=\textwidth]{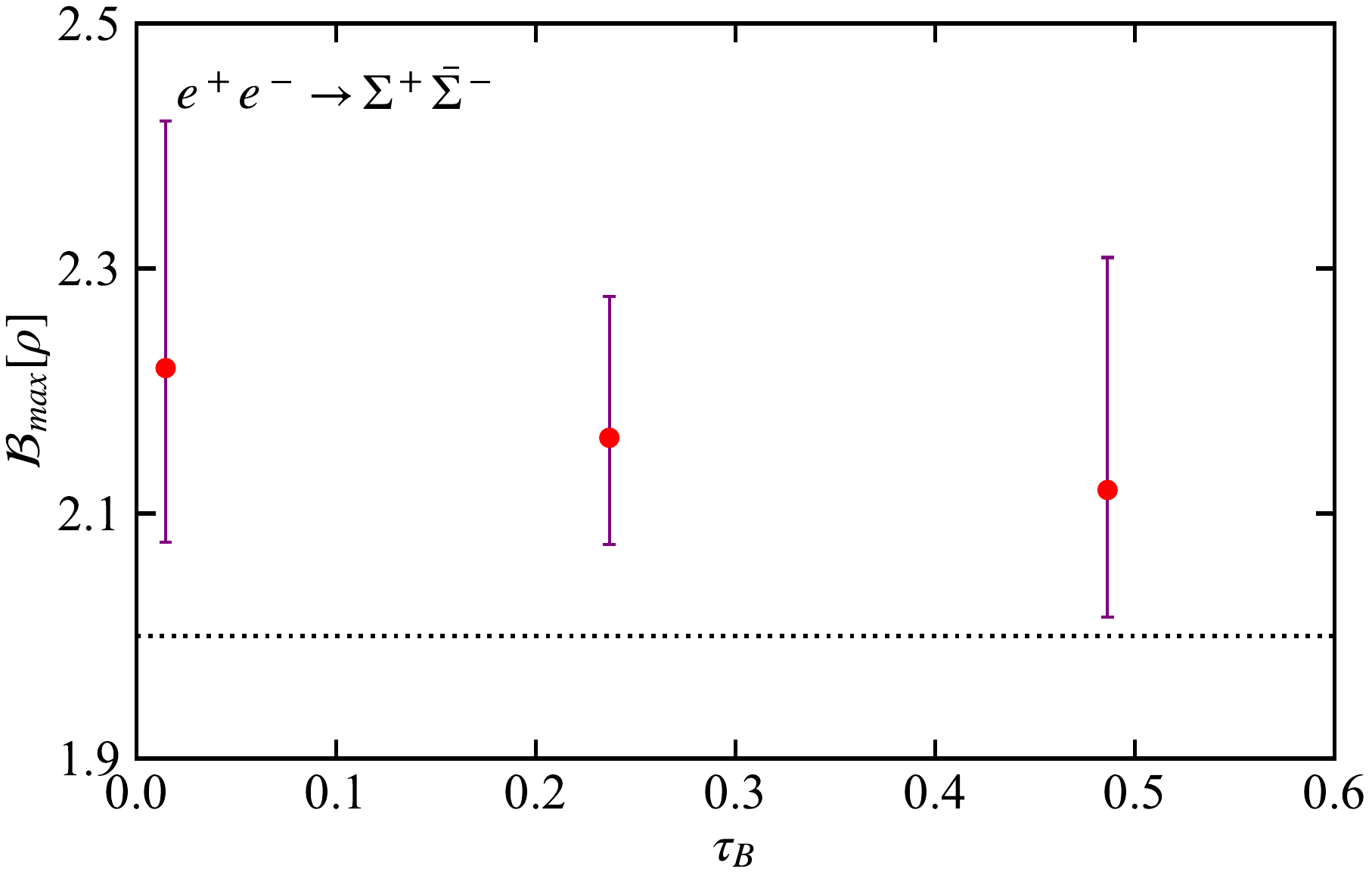}
        \caption{}
    \label{BESIII:d}
    \end{subfigure}
    \hfill
    \begin{subfigure}{0.32\textwidth}
        \includegraphics[width=\textwidth]{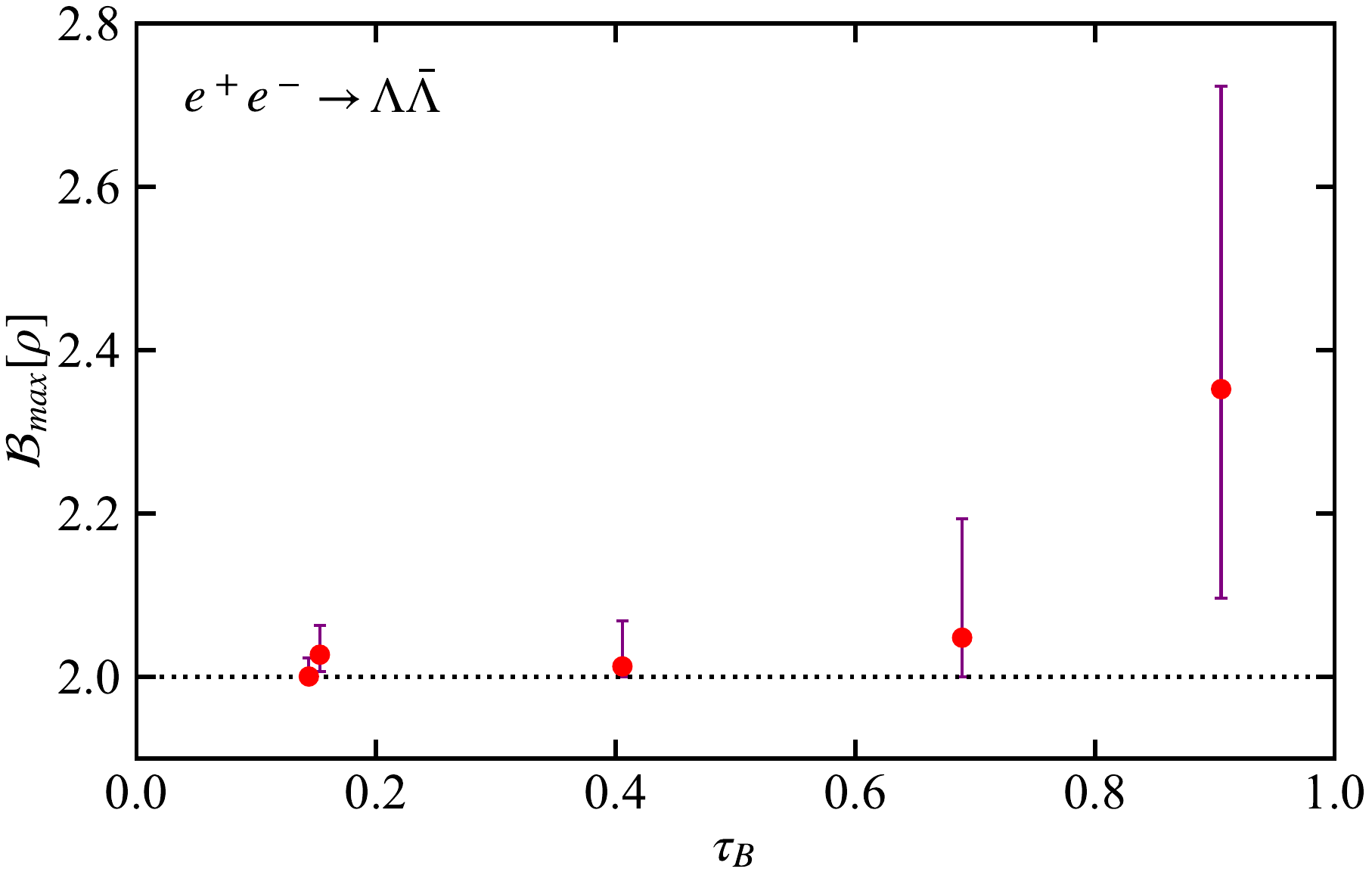}
        \caption{}
    \label{BESIII:e}
    \end{subfigure}
    \hfill
    \begin{subfigure}{0.32\textwidth}
        \includegraphics[width=\textwidth]{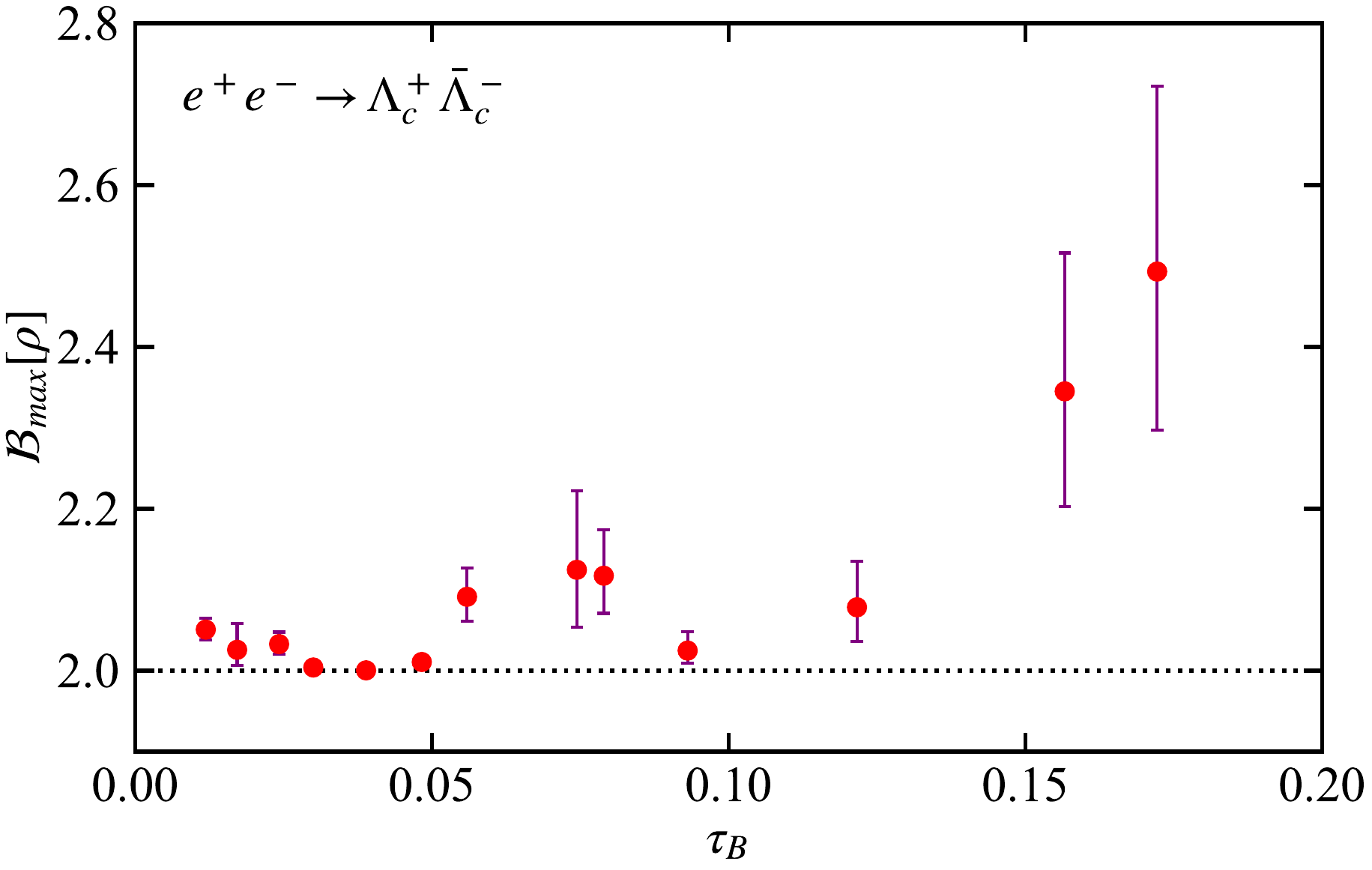}
        \caption{}
    \label{BESIII:f}
    \end{subfigure}
    \vskip\baselineskip
    \centering
    \begin{subfigure}{0.32\textwidth}
        \includegraphics[width=\textwidth]{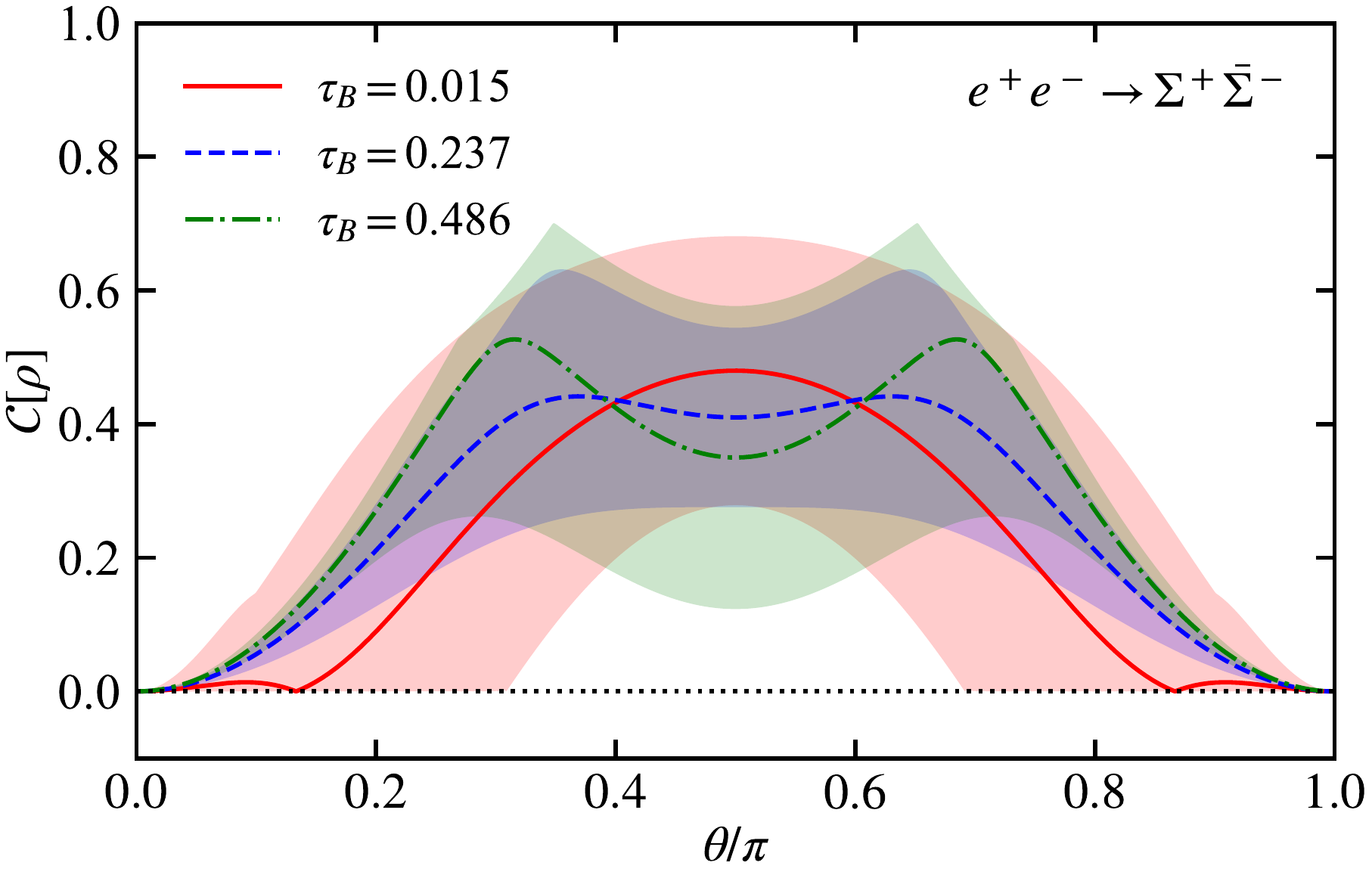}
        \caption{}
    \label{BESIII:g}
    \end{subfigure}
    \hfill
    \begin{subfigure}{0.32\textwidth}
        \includegraphics[width=\textwidth]{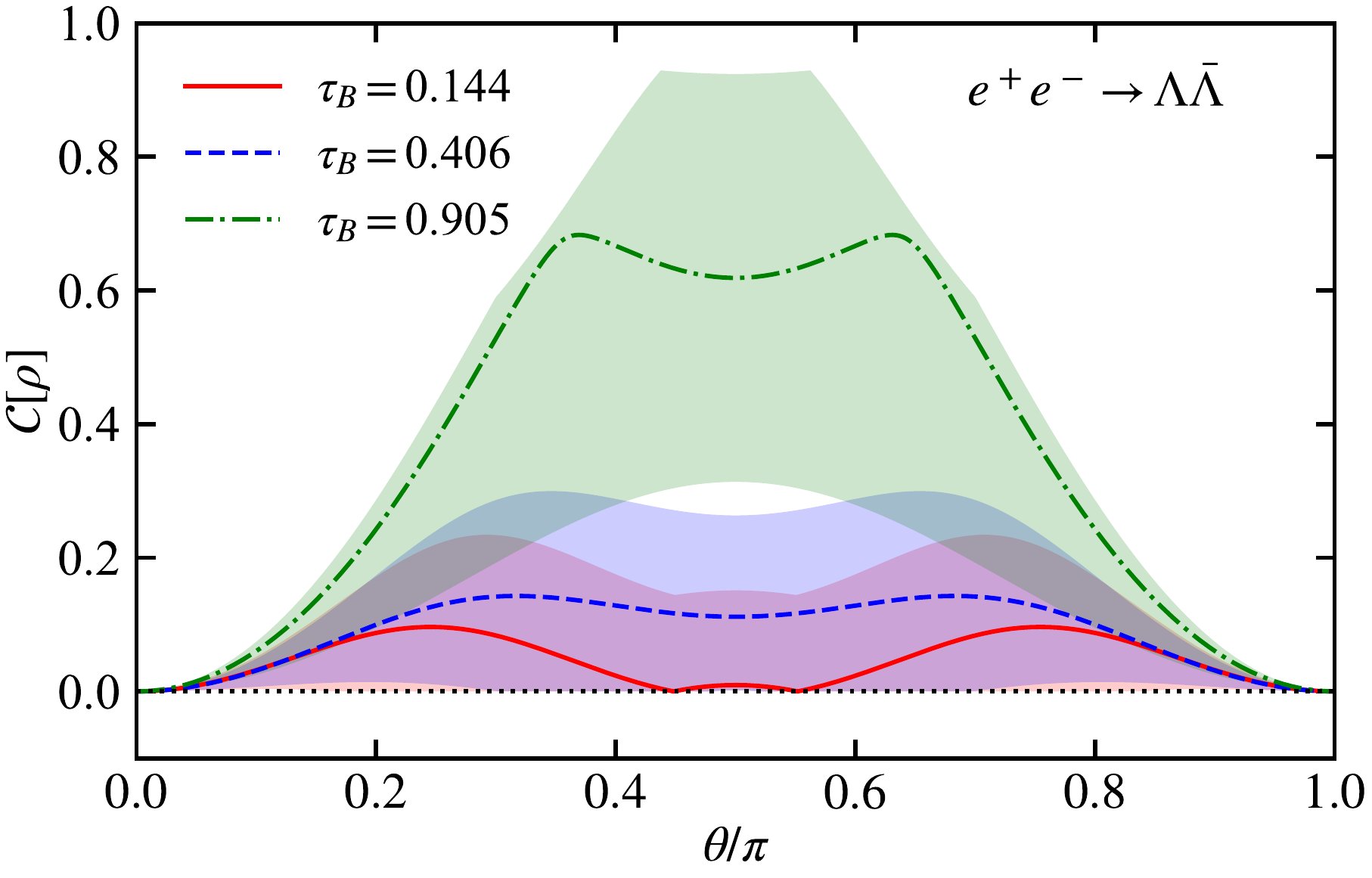}
        \caption{}
    \label{BESIII:h}
    \end{subfigure}
    \hfill
    \begin{subfigure}{0.32\textwidth}
        \includegraphics[width=\textwidth]{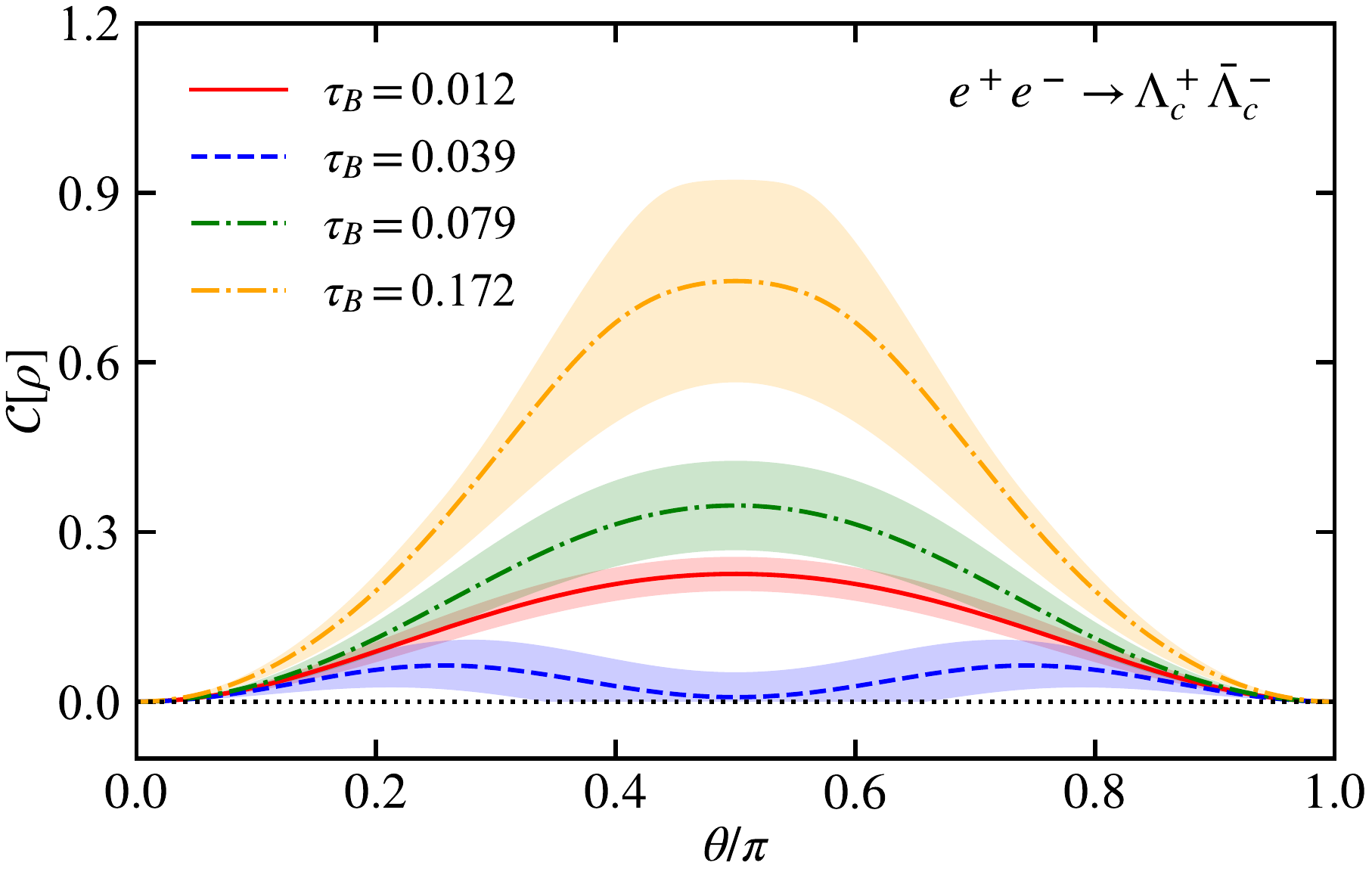}
        \caption{}
    \label{BESIII:i} 
    \end{subfigure}
    \vskip\baselineskip
    \begin{subfigure}{0.32\textwidth}
        \includegraphics[width=\textwidth]{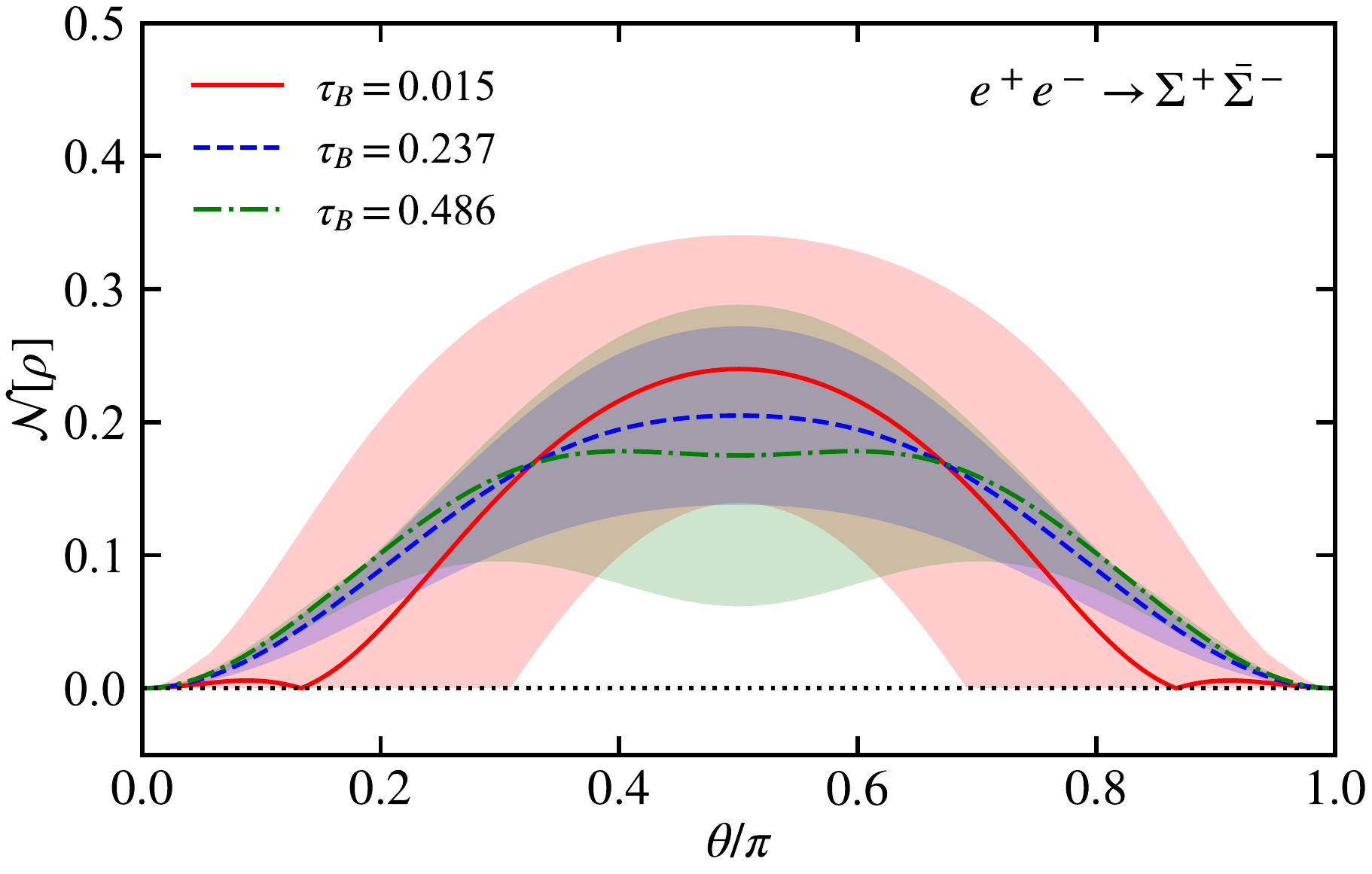}
        \caption{}
    \label{BESIII:j}
    \end{subfigure}
    \hfill
    \begin{subfigure}{0.32\textwidth}
        \includegraphics[width=\textwidth]{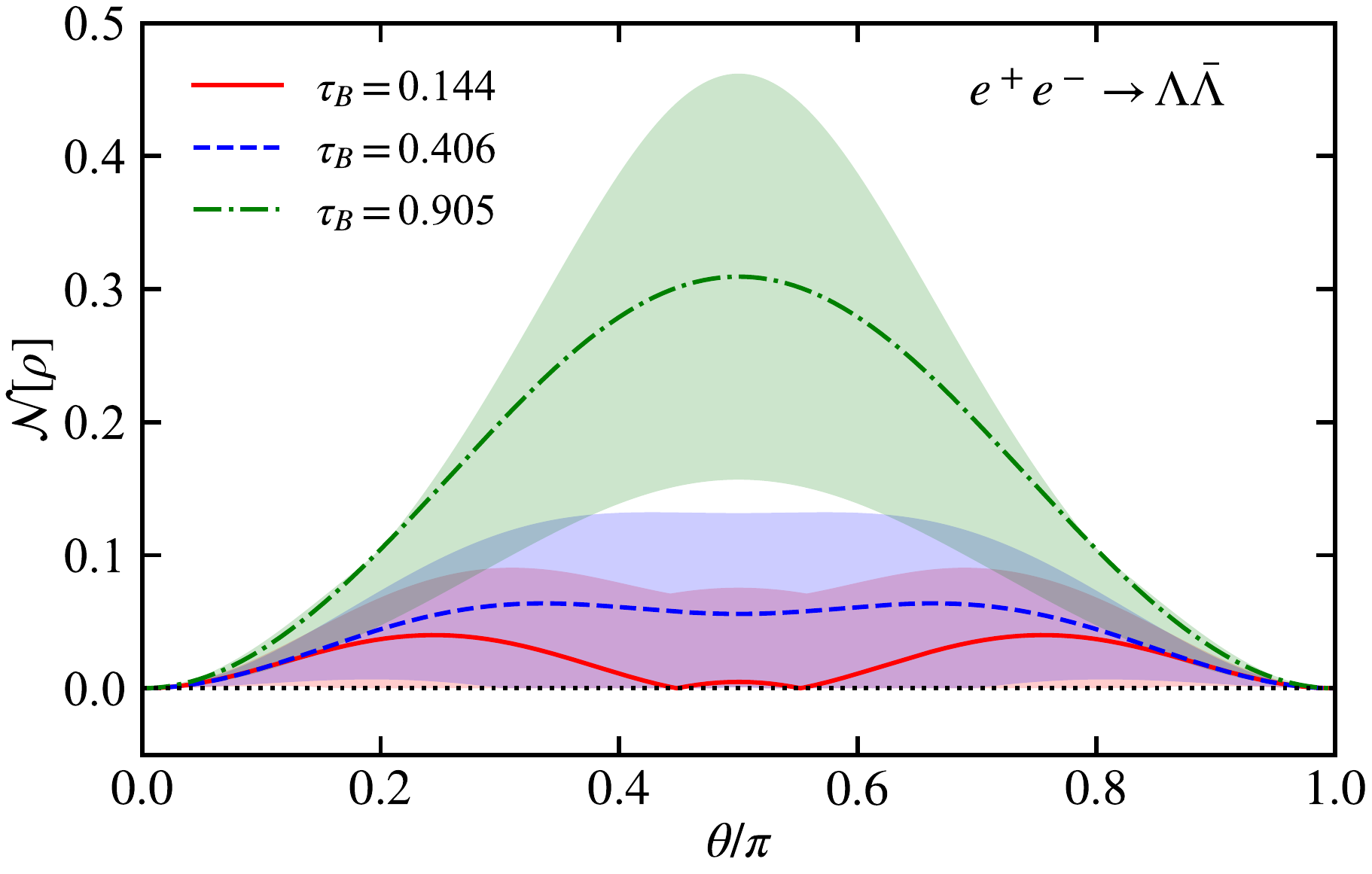}
        \caption{}
    \label{BESIII:k}
    \end{subfigure}
    \hfill
    \begin{subfigure}{0.32\textwidth}
        \includegraphics[width=\textwidth]{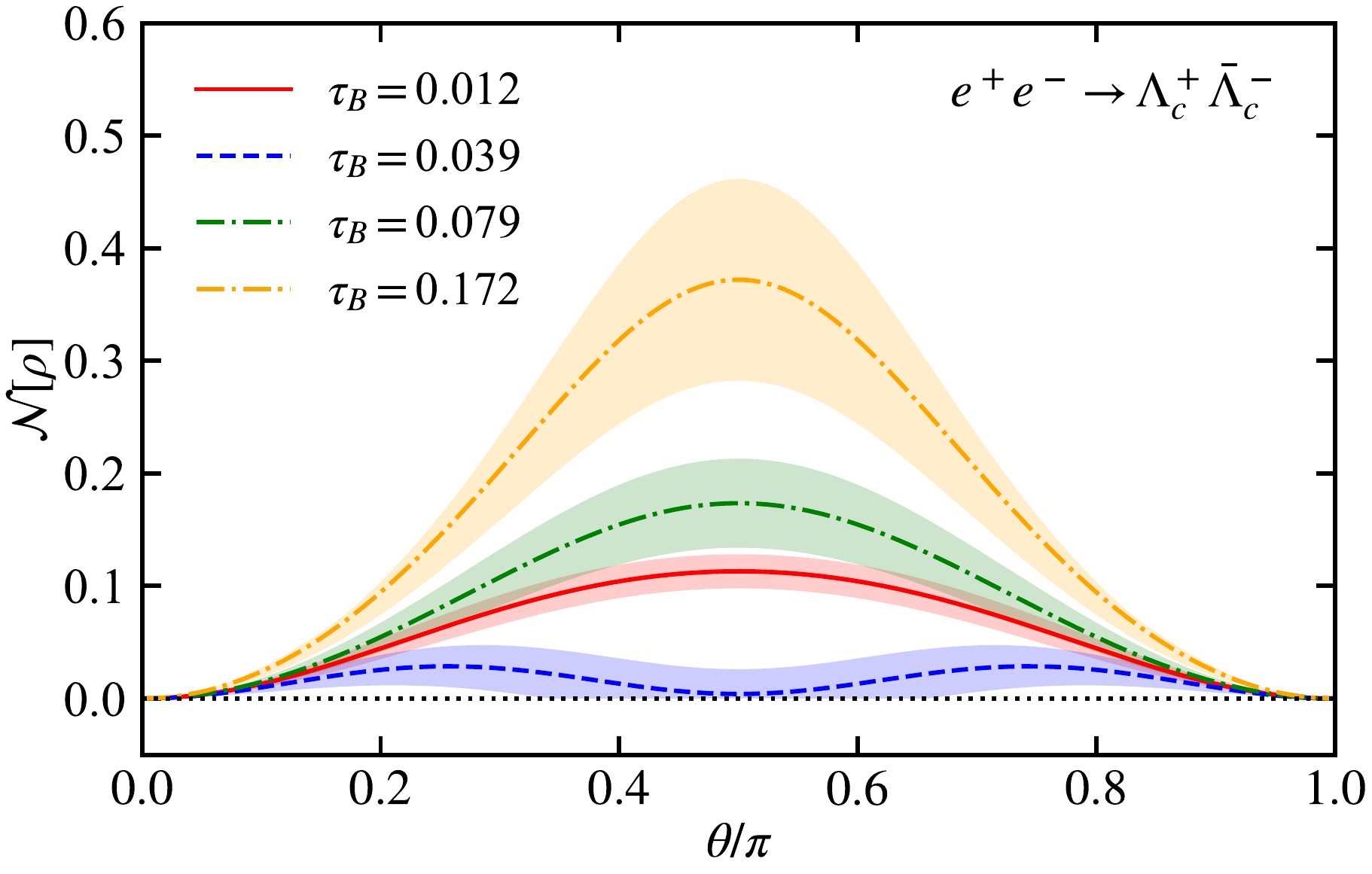}
        \caption{}
    \label{BESIII:l}
    \end{subfigure}
    \vskip\baselineskip
    \begin{subfigure}{0.32\textwidth}
        \includegraphics[width=\textwidth]{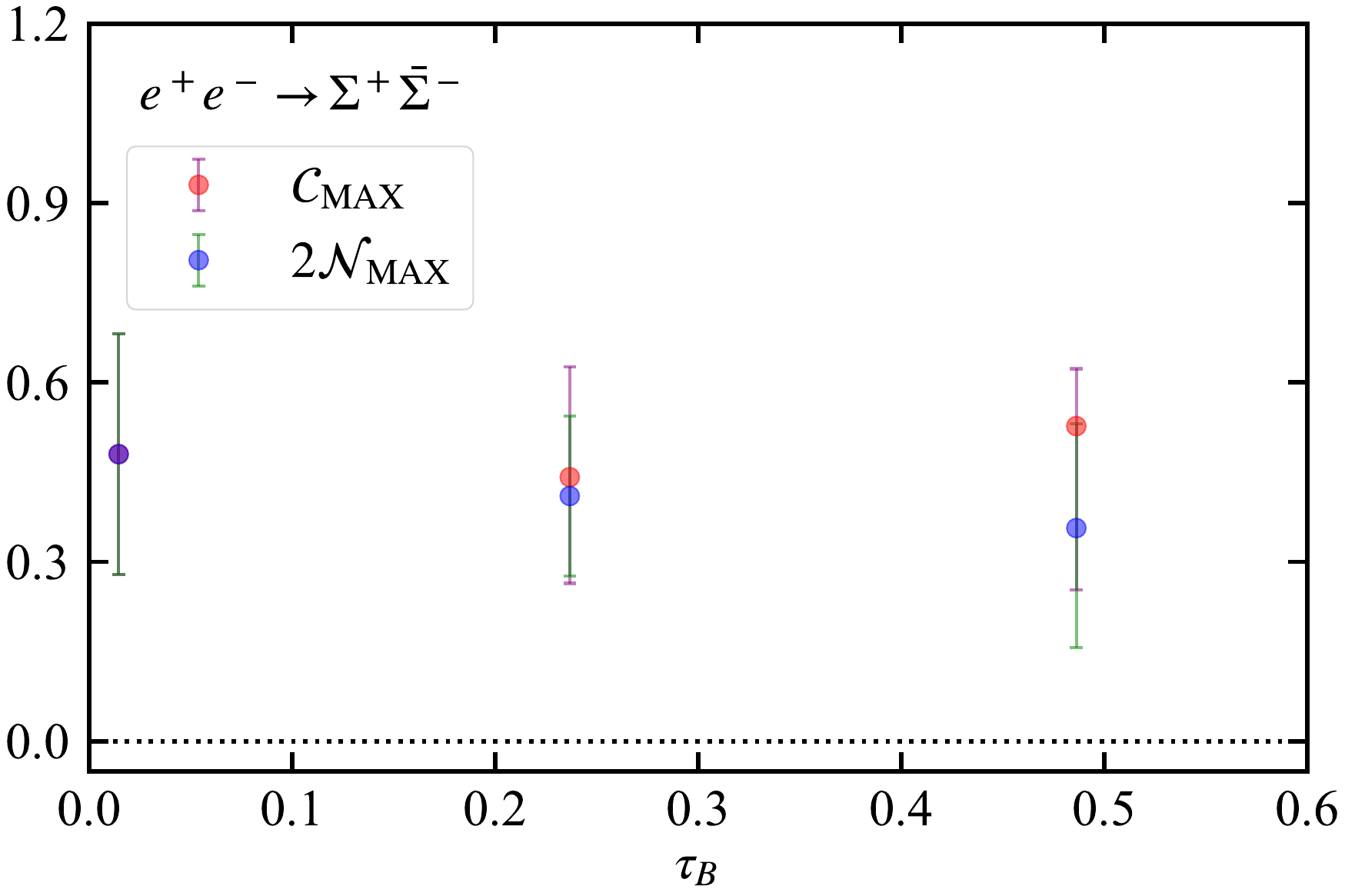}
        \caption{}
    \label{BESIII:m}
    \end{subfigure}
    \hfill
    \begin{subfigure}{0.32\textwidth}
        \includegraphics[width=\textwidth]{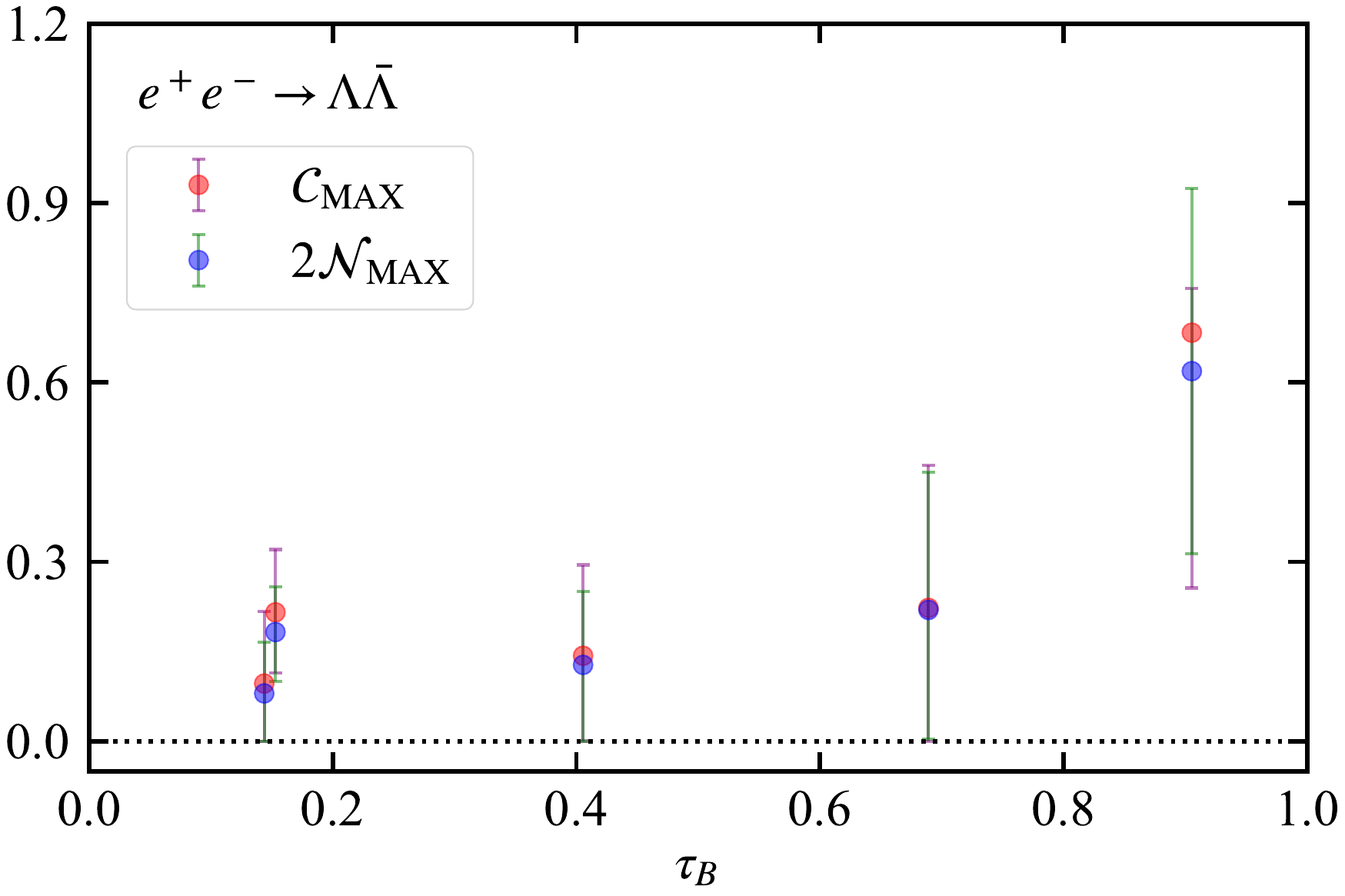}
        \caption{}
    \label{BESIII:n}
    \end{subfigure}
    \hfill
    \begin{subfigure}{0.32\textwidth}
        \includegraphics[width=\textwidth]{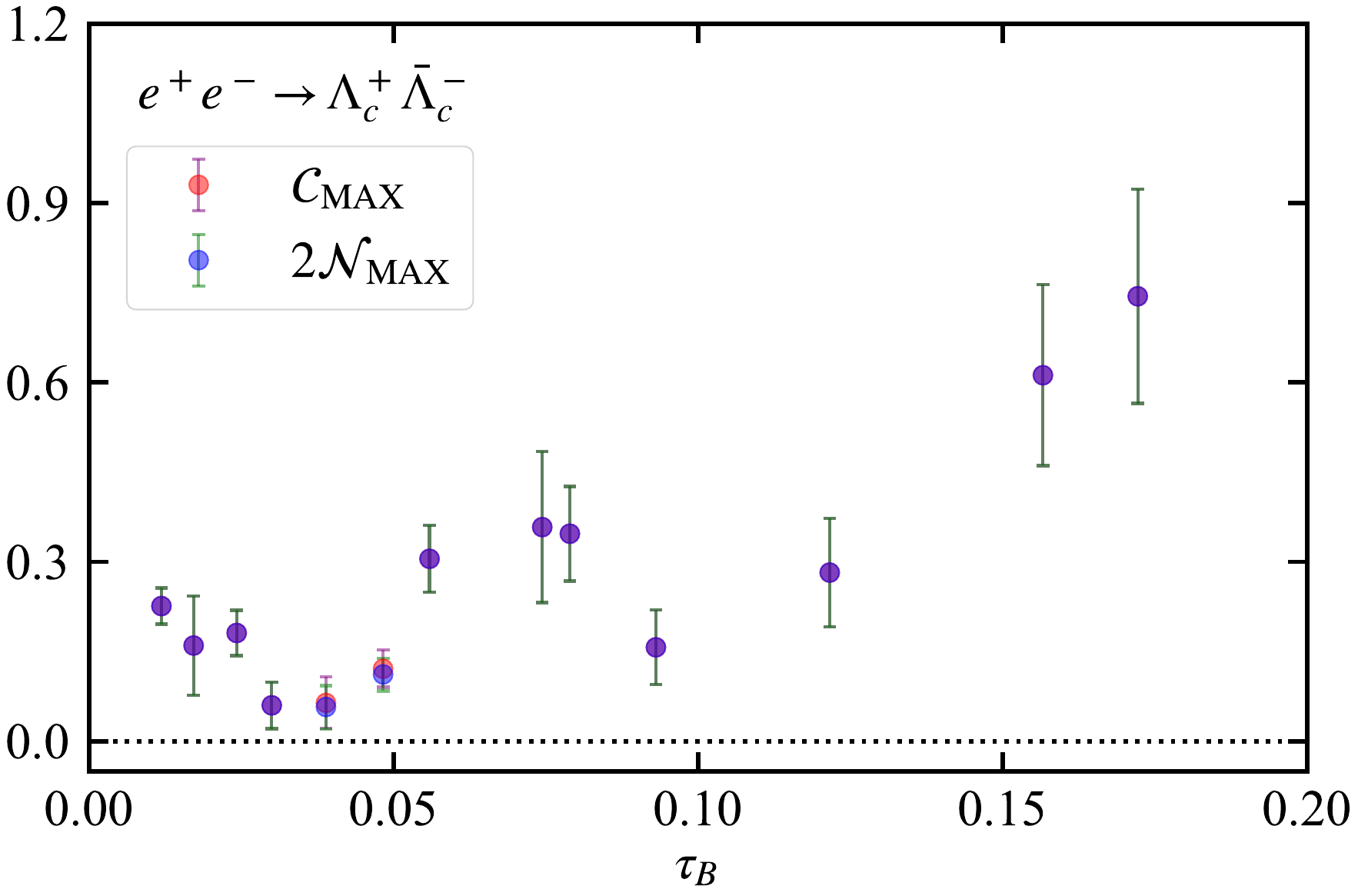}
        \caption{}
    \label{BESIII:o}
    \end{subfigure}
    \caption{Quantum information observables in $B\bar{B}$ systems in $e^+e^-\rightarrow B\bar{B}$ with $B=\Sigma^+$, $\Lambda$, and $\Lambda_c^+$. The data are taken from the BESIII experiment.}
    \label{fig:BESIII}
\end{figure}

Figure~\ref{fig:BESIII} presents the quantum information observables for $B\bar{B}$ with $B=\Sigma^+$, $\Lambda$, and $\Lambda_c^+$ produced in $e^+e^-\rightarrow B\bar{B}$, where the bands represent the errors of these observables. Each column corresponds to one baryon species, while each row corresponds to one observable. 
To facilitate the comparison of observales for different baryon species at different energies, we introduce a dimensionless variable $\tau_B=s/(4M^2)-1$, where $M$ denotes the mass of baryon $B$. 

Figures  (\ref{BESIII:a})-(\ref{BESIII:c}) show the Bell variable $\mathcal{B}$ as functions of the scattering angle $\theta$ for some values of $\tau_B$. The solid black (horizontal) line indicates the threshold value $\mathcal{B}=2$ above which the CHSH inequality is violated. Figures  (\ref{BESIII:d})-(\ref{BESIII:f}) display the maximum values of $\mathcal{B}$ at specific $\theta$ versus $\tau_B$. The concurrence $\mathcal{C}$, a measure of entanglement, is plotted versus $\theta$ for some values of $\tau_B$ in Figs. (\ref{BESIII:g})-(\ref{BESIII:i}), with the dotted (horizontal) line representing the separability bound $\mathcal{C}=0$. Similarly, Figs. (\ref{BESIII:j})-(\ref{BESIII:l}) present the negativity $\mathcal{N}$ versus $\theta$ for some values of $\tau_B$, with the bound $\mathcal{N}=0$ shown in the dotted (horizontal) line. Finally, Figs.  (\ref{BESIII:m})-(\ref{BESIII:o}) show the maximum values of $\mathcal{C}$ and $2\mathcal{N}$ versus $\tau_B$, along with the bounds $\mathcal{C}=0$ and $2\mathcal{N}=0$ indicated by the dotted (horizontal) line. For the spin state of the $B\bar{B}$ system, the CHSH inequality is violated when $\mathcal{B}>2$, and the state is entangled iff $\mathcal{C}>0$ or $\mathcal{N}>0$. 

As shown in Fig.~\ref{fig:BESIII}, $\mathcal{B}$, $\mathcal{C}$ and $\mathcal{N}$ are all symmetric functions of $\theta$ relative to $\theta=\pi/2$ in the range $\theta\in[0,\pi]$. 
The Bell variable $\mathcal{B}$ always reaches its maximum value $\mathcal{B}_\text{max}=2\sqrt{1+\alpha^2}$ at $\theta=\pi/2$. There is always a range of the scattering angle $\theta\in[0,\theta_0]\cup [\pi-\theta_0,\pi]$ in which the CHSH inequality is satisfied, which is different from the result observed in elementary particle-antiparticle systems. Even when experimental uncertainties are taken into account, both $\mathcal{C}$ and $\mathcal{N}$ remain greater than zero in most cases, indicating that the spin states of the hyperon and its antihyperon are almost always entangled. However, according to our results, there is no clear relationship among $\mathcal{B}$, $\mathcal{C}$ and $\mathcal{N}$ as functions of $\theta$ and $\sqrt{s}$, so are their maximum values. 

\subsection{Discussions} \label{Sec:Discussions}

In this subsection, we discuss quantum information observables for the hyperon-antihyperon systems and their implications for the experimental measurement of the parton entanglement in heavy-flavored mesons.

\subsubsection{Relations between quantum information observables}

We look for the relationship among the Bell variable, concurrence and negativity that characterize quantum correlations.

For an arbitrary density operator $\rho$ for a two-qubit system, we have the following properties for the Bell variable, concurrence and negativity~\cite{Verstraete:2001skx,Audenaert:2000htf,Miranowicz:2004sre}
\begin{equation}
\mathcal{B}[\rho]\leq2\sqrt{1+\mathcal{C}^2[\rho]},
\label{eqa:BC-relation}
\end{equation}
and
\begin{equation}
F(\mathcal{C}[\rho])\equiv \sqrt{(1-\mathcal{C}[\rho])^2+\mathcal{C}^2[\rho]}-(1-\mathcal{C}[\rho])\leq 2\mathcal{N}[\rho]\leq\mathcal{C}[\rho]\,.
\label{eqa:CN-relation}
\end{equation}
Equation (\ref{eqa:BC-relation}) provides the relation between the Bell nonlocality and entanglement in the hierarchy of quantumness 
\begin{equation}
\mathrm{Bell\;\; nonlocality}\; \subset\; \mathrm{Entanglement}\;,
\end{equation}
which means that any nonlocal state must be entangled, but not all entangled states are nonlocal or some entangled states are local. 
This conclusion is also consistent with our discussion about Bell nonlocality in Section~\ref{sec:Bell-nonlocality}.
Equation (\ref{eqa:CN-relation}) is satisfied since the concurrence $\mathcal{C}$ is a complete and monotonous measure of the entanglement in a two-qubit system, corresponding to the formation of entanglement $E_F$~\cite{Hill:1997pfa}. 
Therefore, $\mathcal{C}$ represents the total cost of entanglement.
However, the physical meaning of the negativity is the ``degree of violation" of an entangled state with respect to the partial transposition operation, which is different from the concurrence. 
Furthermore, there exists an ordering discrepancy between them~\cite{Miranowicz:2004sre}: for two different systems with $\rho_1$ and $\rho_2$, it is possible to have $\mathcal{C}[\rho_1]>\mathcal{C}[\rho_2]$ but $\mathcal{N}[\rho_1]<\mathcal{N}[\rho_2]$. 
Nonetheless, the negativity remains a valid quantifier of the entanglement particularly in high-dimensional systems, where the concurrence is no longer well-defined and the entanglement of formation $E_F$ becomes intractable in computation. In such cases, the negativity serves as an optimal choice for characterizing the entanglement.

\begin{figure}[p]
    \centering
    \begin{subfigure}{0.4\textwidth}
        \includegraphics[width=\textwidth]{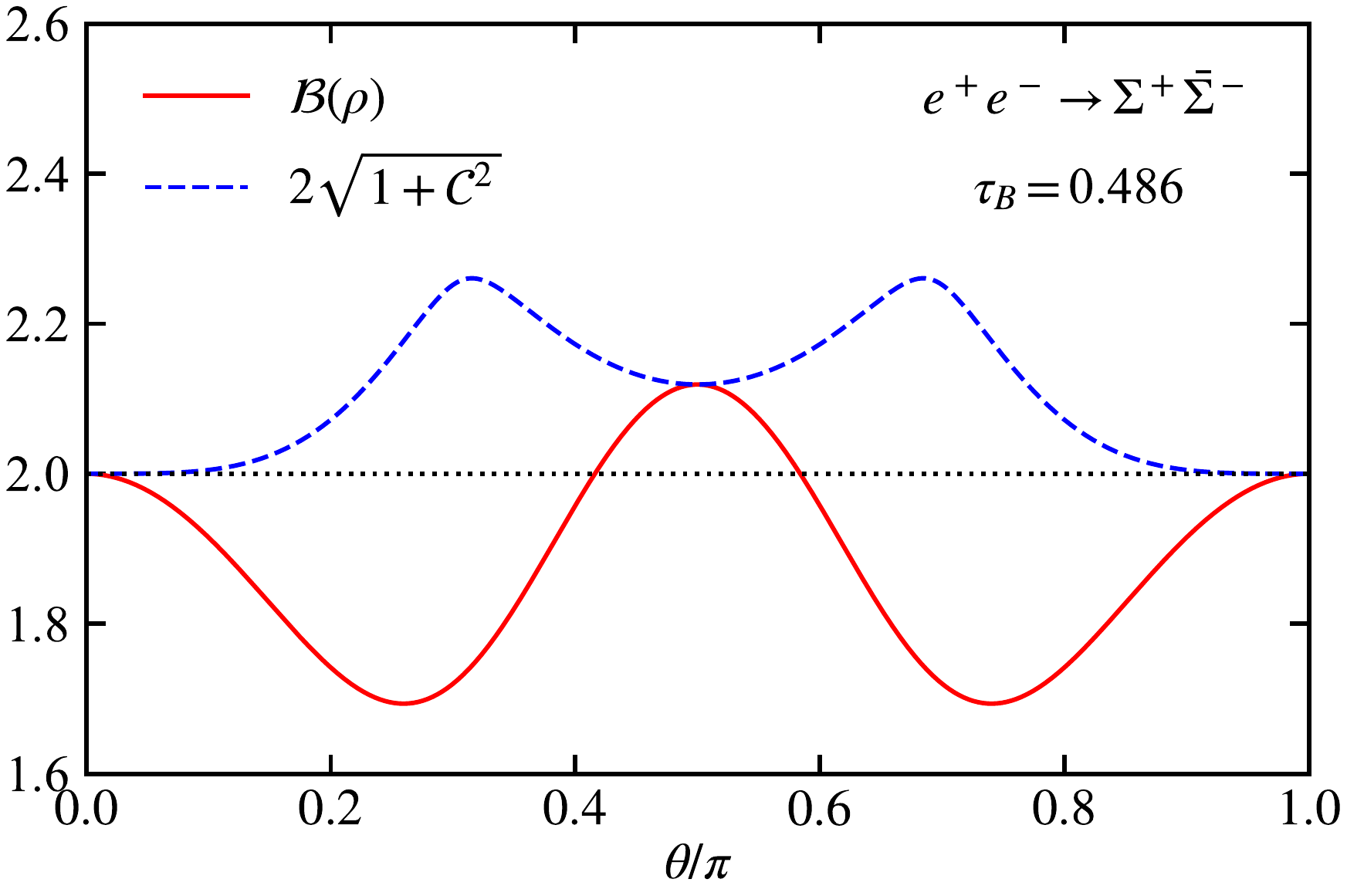}
        \caption{}
    \label{comparasion:a}
    \end{subfigure}
    \hspace{0.05\textwidth}
    \begin{subfigure}{0.4\textwidth}
        \includegraphics[width=\textwidth]{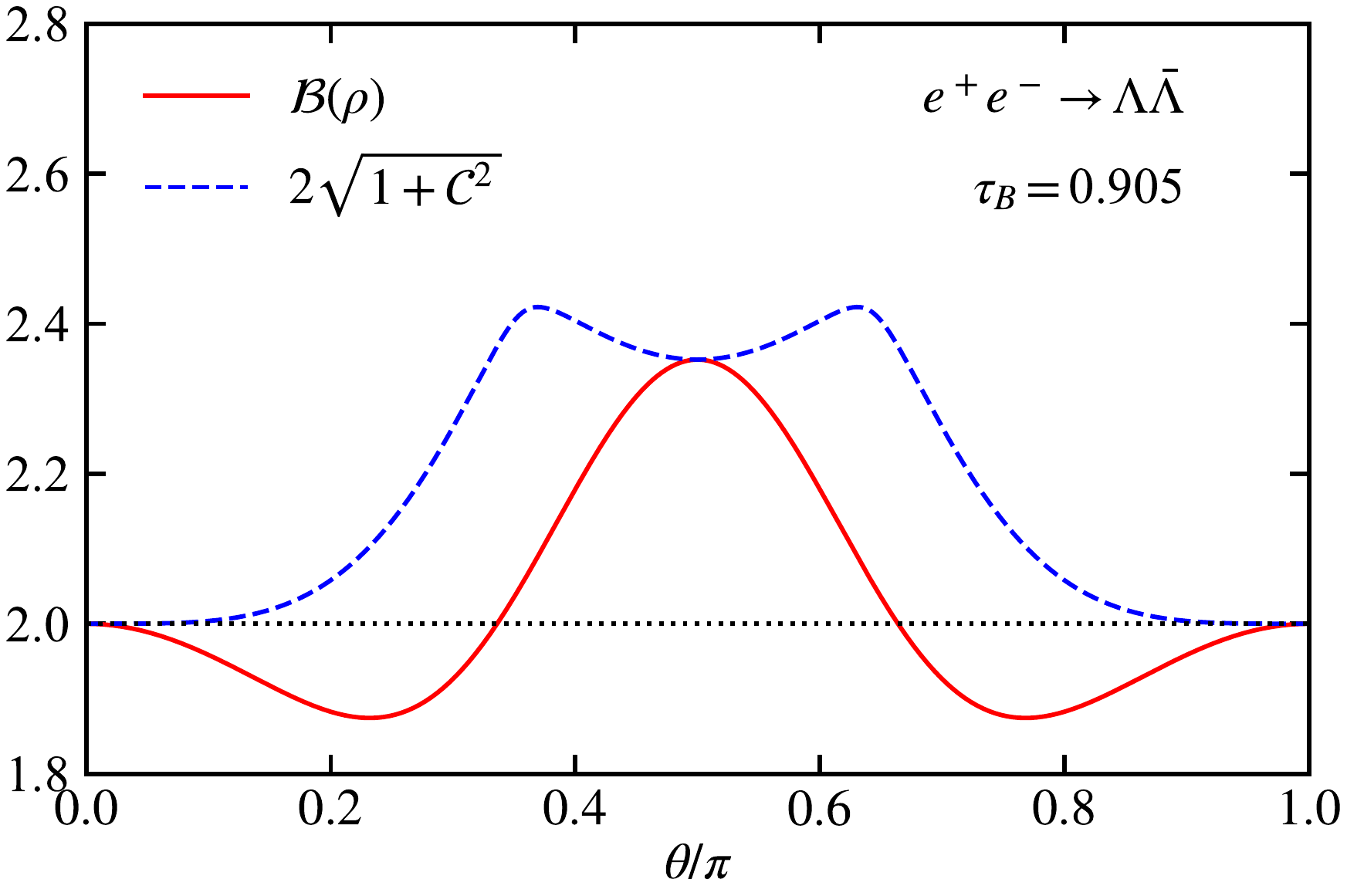}
        \caption{}
    \label{comparasion:b}
    \end{subfigure}
    \vskip\baselineskip
    \begin{subfigure}[b]{0.4\textwidth}
        \includegraphics[width=\textwidth]{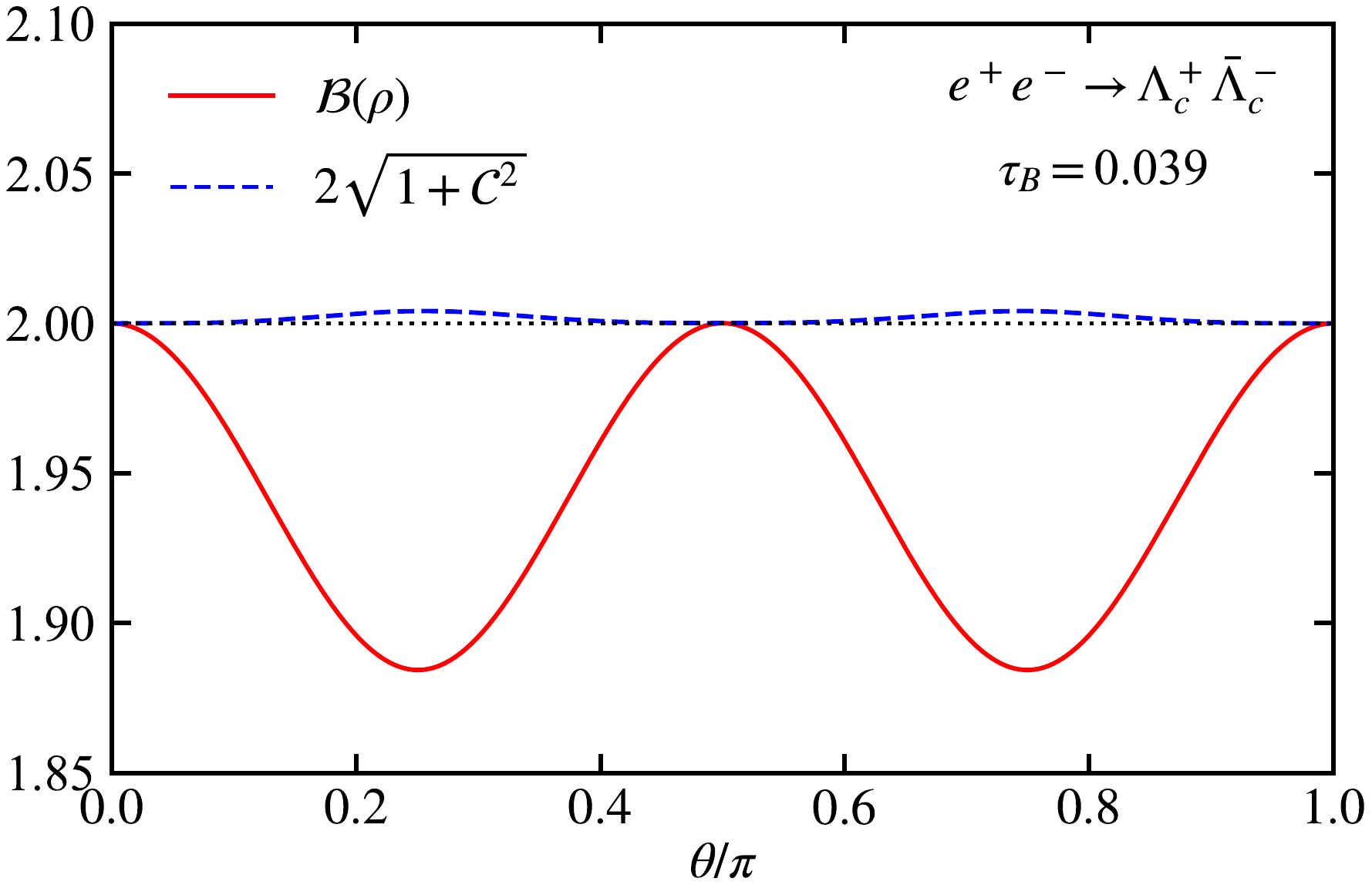}
        \caption{}
    \label{comparasion:c}
    \end{subfigure}
    \hspace{0.05\textwidth}
    \begin{subfigure}[b]{0.4\textwidth}
        \includegraphics[width=\textwidth]{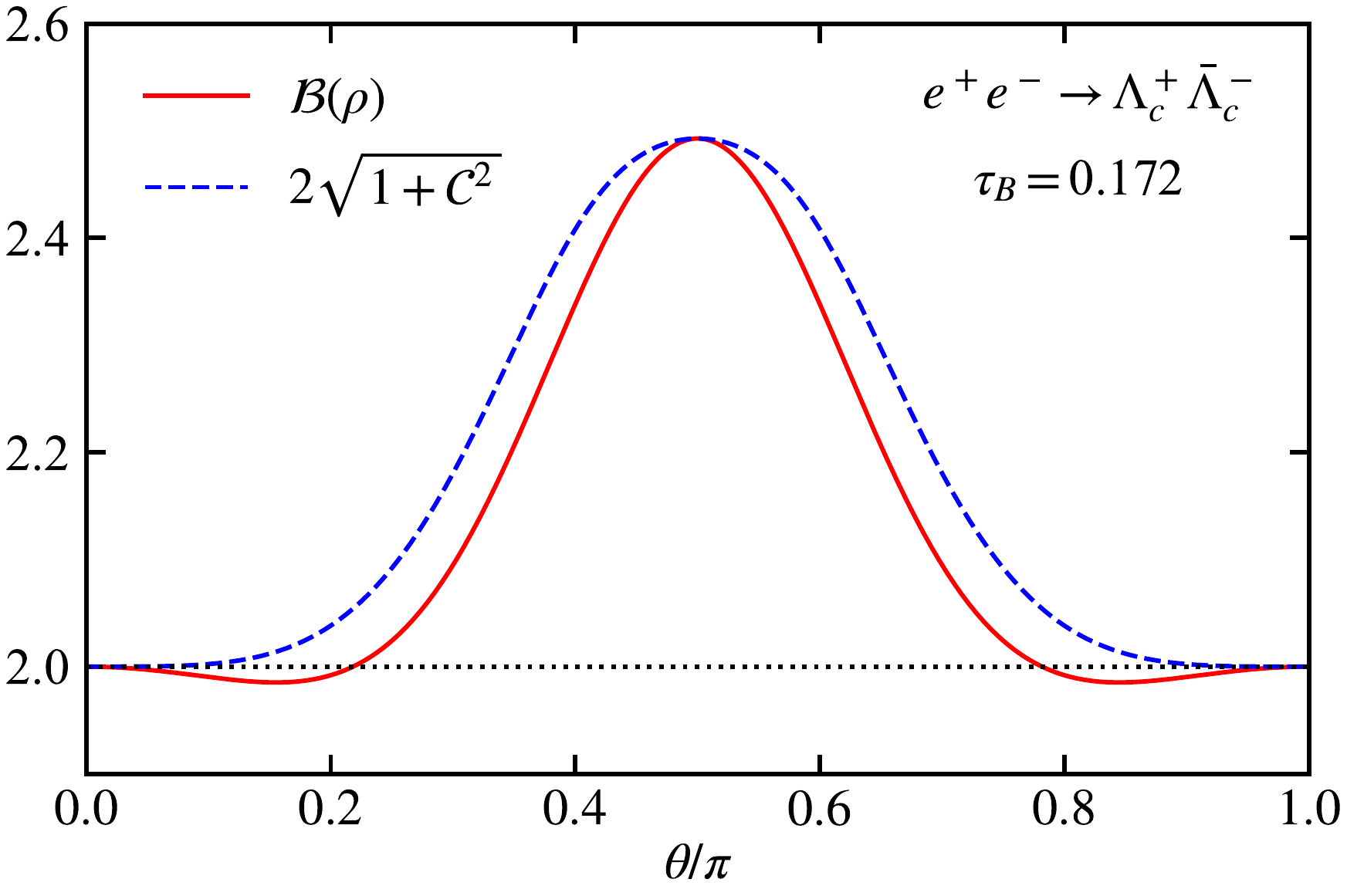}
        \caption{}
    \label{comparasion:d}
    \end{subfigure}
    \vskip\baselineskip
    \begin{subfigure}{0.4\textwidth}
        \includegraphics[width=\textwidth]{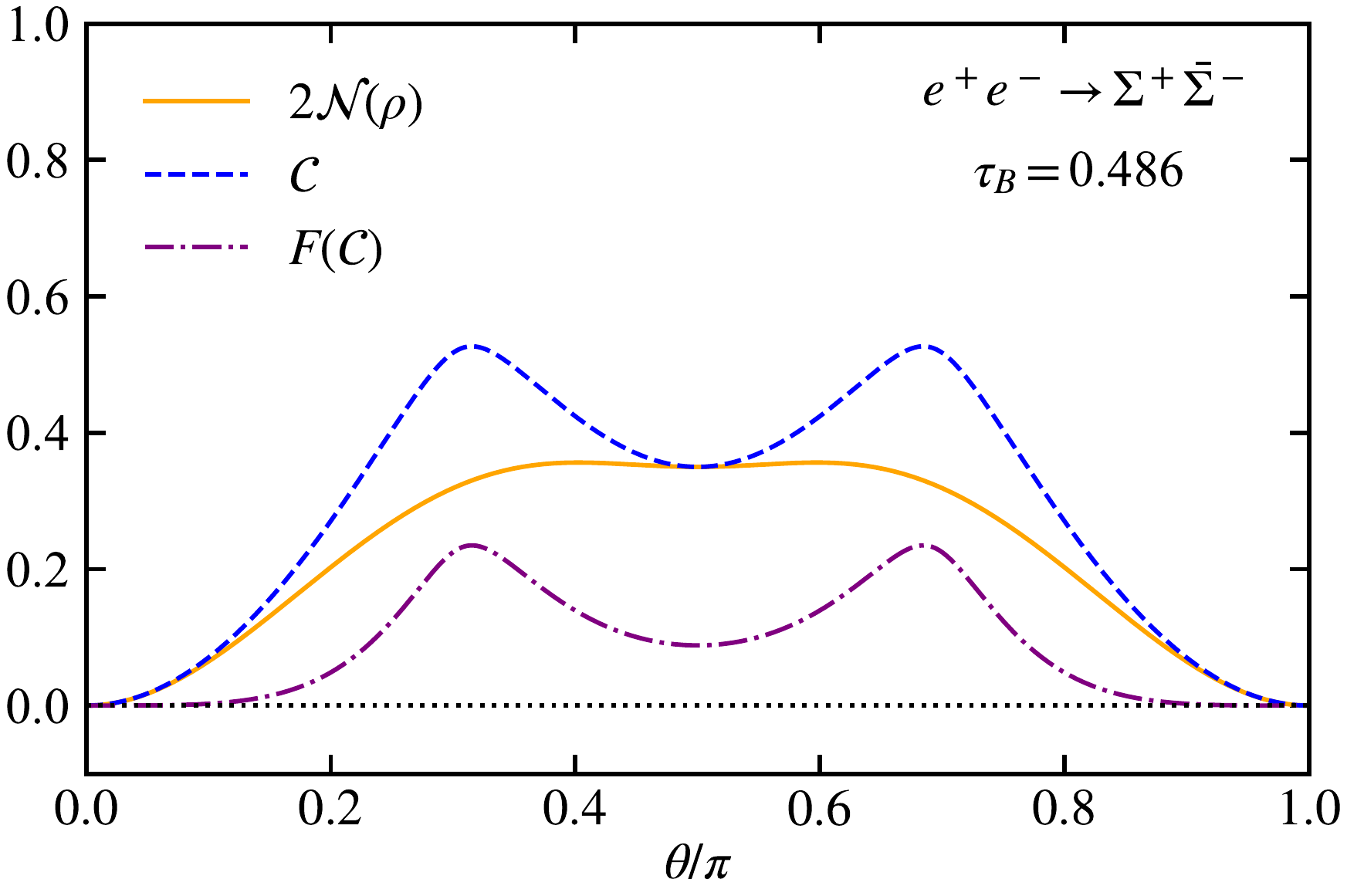}
        \caption{}
    \label{comparasion:e}
    \end{subfigure}
    \hspace{0.05\textwidth}
    \begin{subfigure}{0.4\textwidth}
        \includegraphics[width=\textwidth]{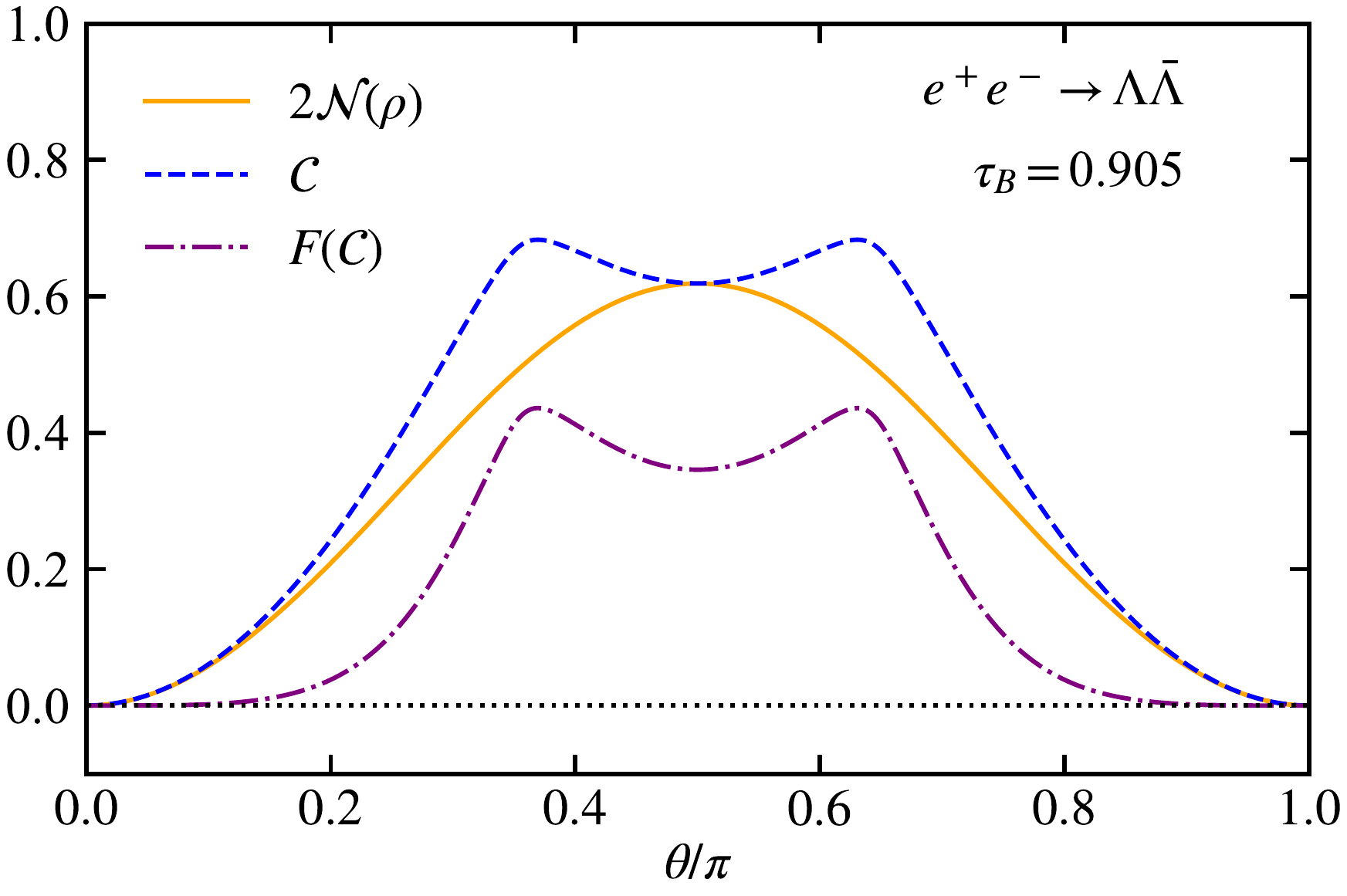}
        \caption{}
    \label{comparasion:f}
    \end{subfigure}
    \vskip\baselineskip
    \begin{subfigure}[b]{0.4\textwidth}
        \includegraphics[width=\textwidth]{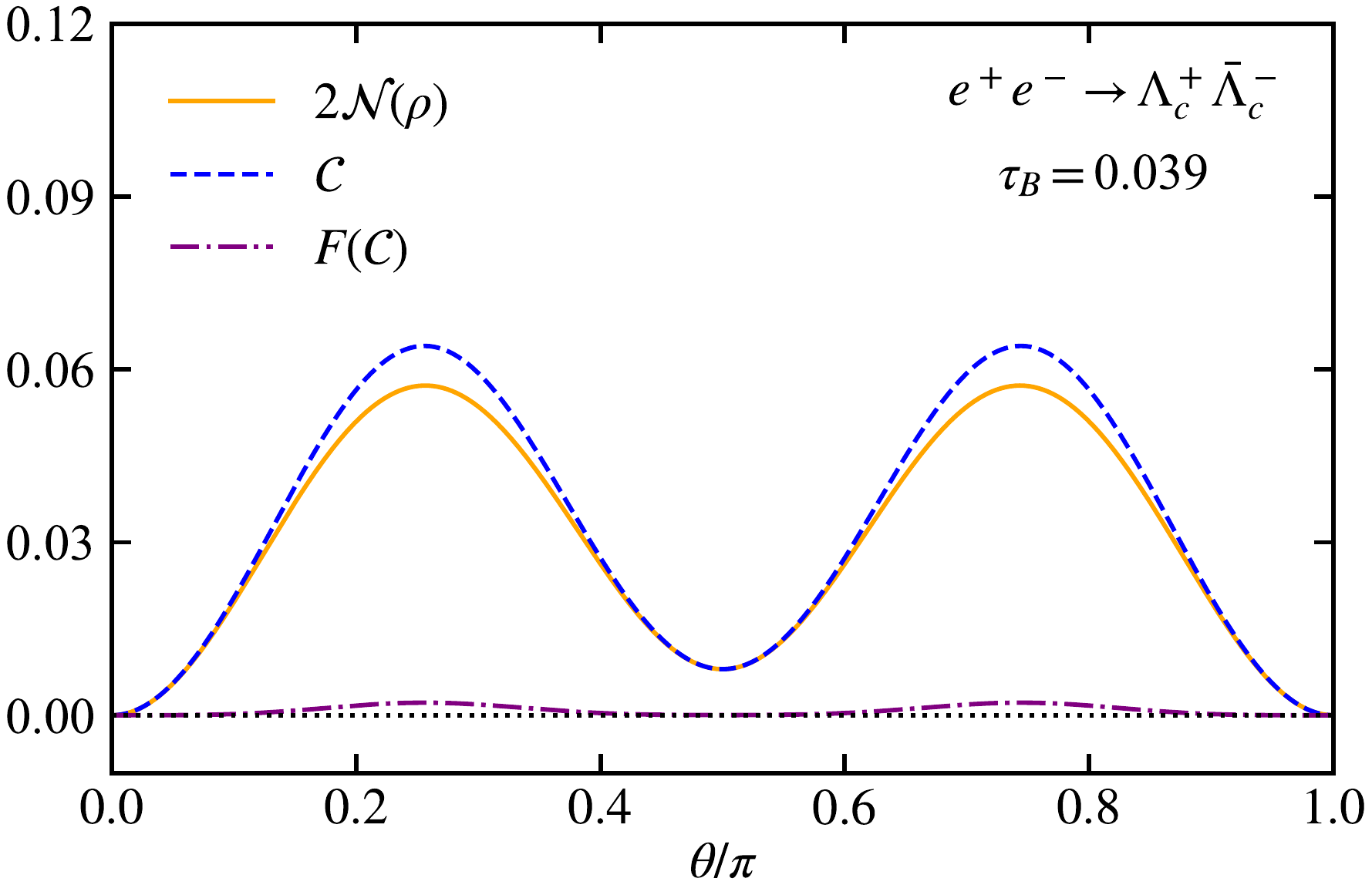}
        \caption{}
    \label{comparasion:g}
    \end{subfigure}
    \hspace{0.05\textwidth}
    \begin{subfigure}[b]{0.4\textwidth}
        \includegraphics[width=\textwidth]{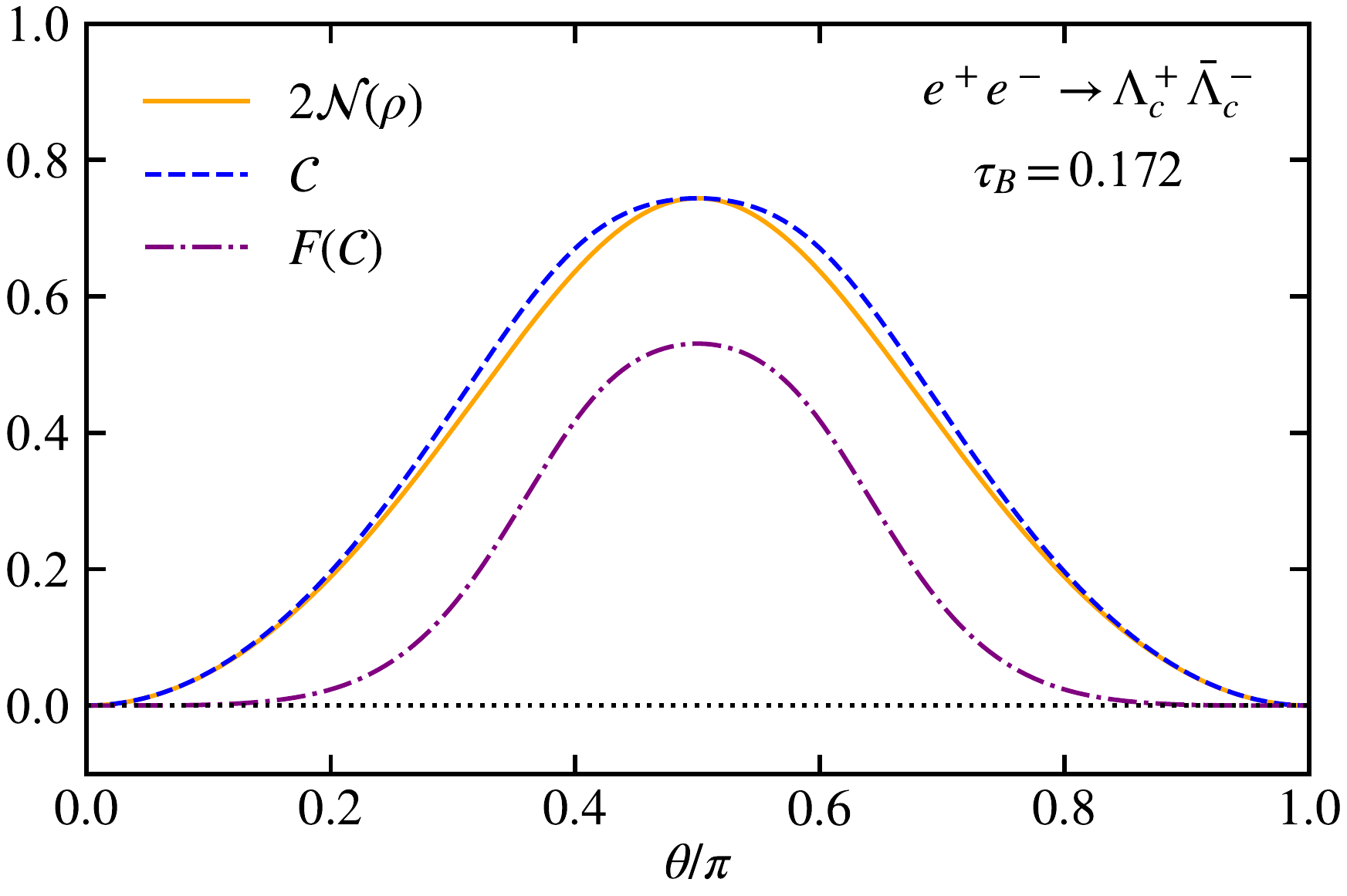}
        \caption{}
    \label{comparasion:h}
    \end{subfigure}
    \caption{The comparison of different quantum information observables as functions of the scattering angle $\theta$ at some collision energy $\sqrt{s}$ in $e^+e^-\rightarrow B\bar{B}$ with $B=\Sigma^+$, $\Lambda$, and $\Lambda_c^+$. }
    \label{fig:quantum-comparasion}
\end{figure}

Figure~\ref{fig:quantum-comparasion} presents the comparison of different quantum information observables as functions of the scattering angle $\theta$ for $B\bar{B}$ produced in $e^+e^-$ annihilation with $B=\Sigma^+$, $\Lambda$, and $\Lambda_c^+$. 
Figs.  (\ref{comparasion:a})-(\ref{comparasion:d}) show the comparison between $\mathcal{B}$ and $2\sqrt{1+\mathcal{C}^2}$ at some collision energy $\sqrt{s}$. The dotted black (horizontal) lines indicate the bound value 2. Figs.  (\ref{comparasion:e})-(\ref{comparasion:h}) show the comparison between $\mathcal{C}$, $F(\mathcal{C})$ and $2\mathcal{N}$ at some $\sqrt{s}$. The dotted black (horizontal) lines indicate the bound value 0. 

The results in Figure~\ref{fig:quantum-comparasion} are all consistent with equalities (\ref{eqa:BC-relation}) and (\ref{eqa:CN-relation}). 
Even if hyperon-antihyperon pairs are entangled in the full range of the scattering angle except $\theta=0,\pi$, the violation of CHSH inequality only appears within a certain range of $\theta$, fully consistent with our previous discussion on the relationship between the Bell nonlocality and entanglement. 
Furthermore, the difference between the concurrence and twice the negativity is not significant, although they exhibit an ordering discrepancy: both measures provide a good description of the entanglement. 

\begin{figure}[p]
    \centering
    \begin{subfigure}{0.3\textwidth}
        \includegraphics[width=\textwidth]{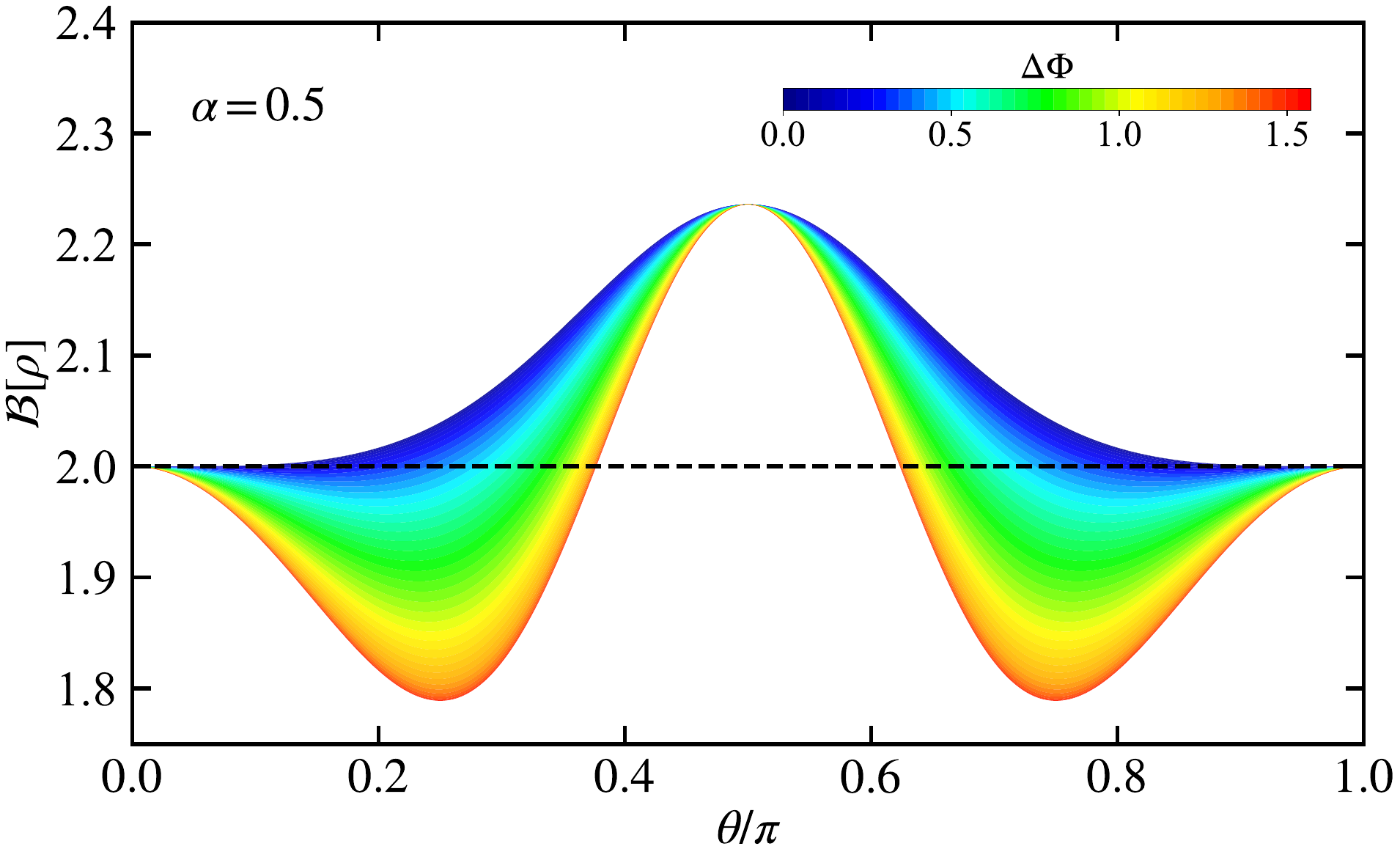}
        \caption{}
    \label{general:a}
    \end{subfigure}
    \hspace{0.027\textwidth}
    \begin{subfigure}{0.3\textwidth}
        \includegraphics[width=\textwidth]{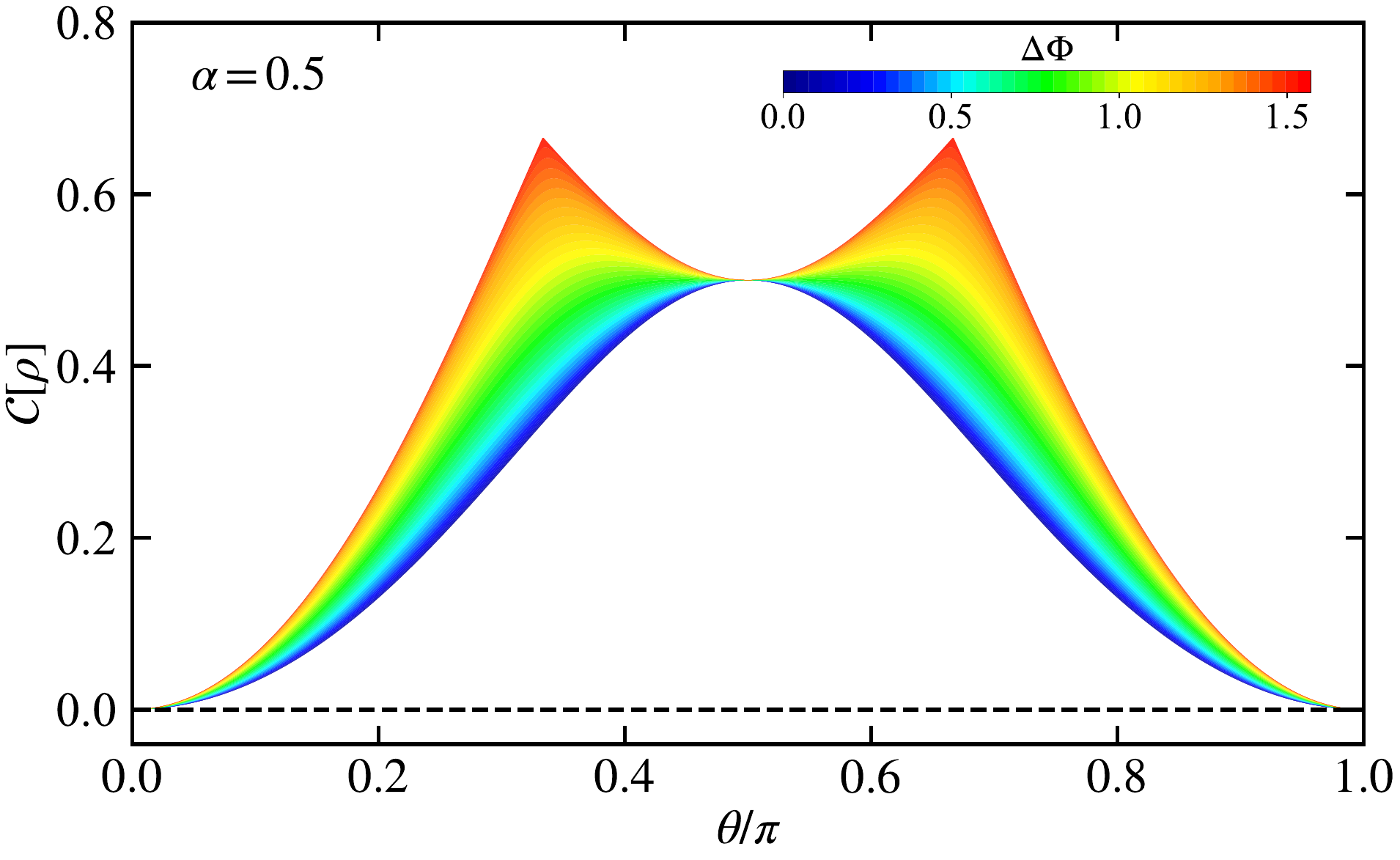}
        \caption{}
    \label{general:b}
    \end{subfigure}
    \hspace{0.027\textwidth}
    \begin{subfigure}[b]{0.3\textwidth}
        \includegraphics[width=\textwidth]{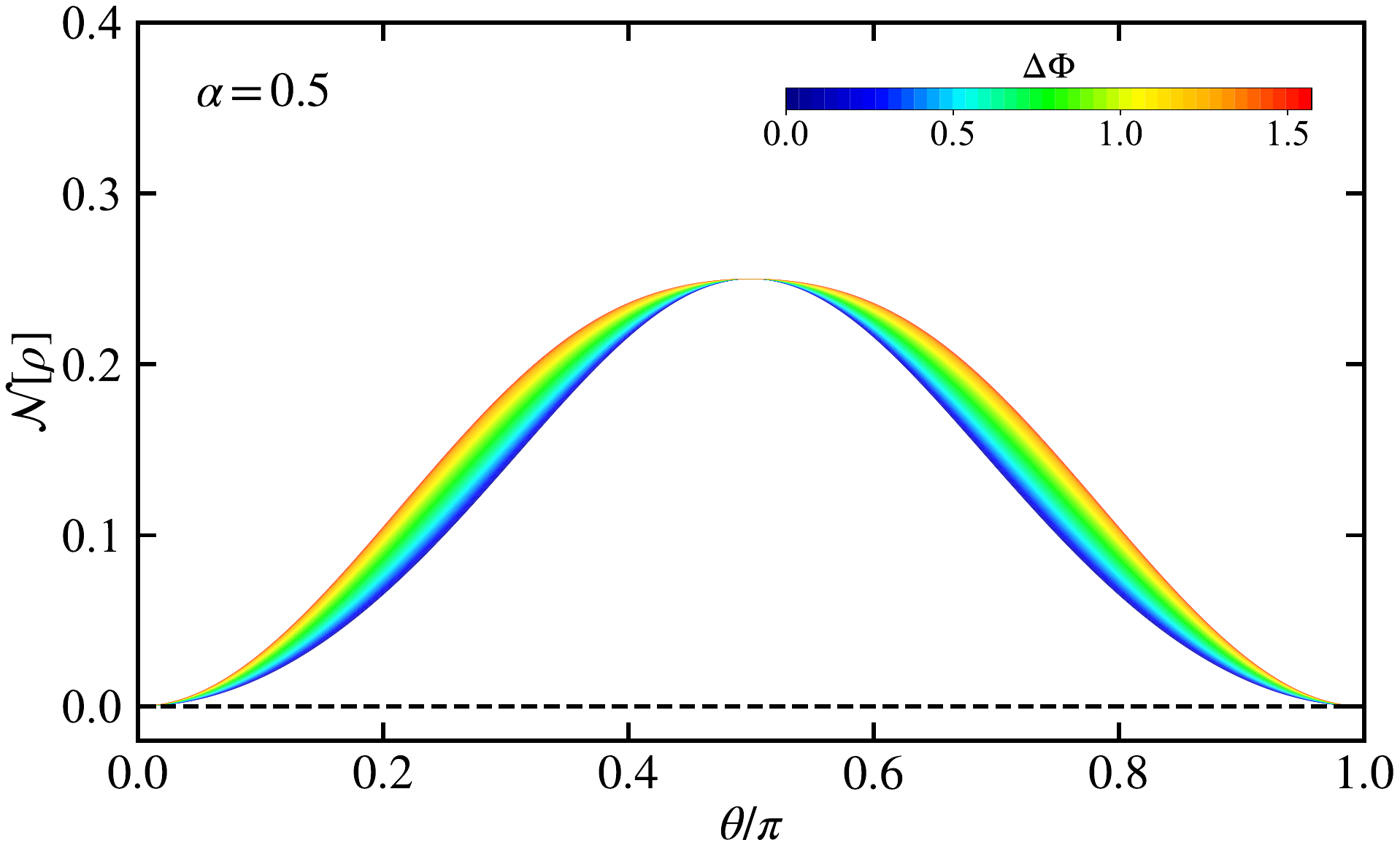}
        \caption{}
    \label{general:c}
    \end{subfigure}
    \vspace{0.2cm}
    \begin{subfigure}[b]{0.3\textwidth}
        \includegraphics[width=\textwidth]{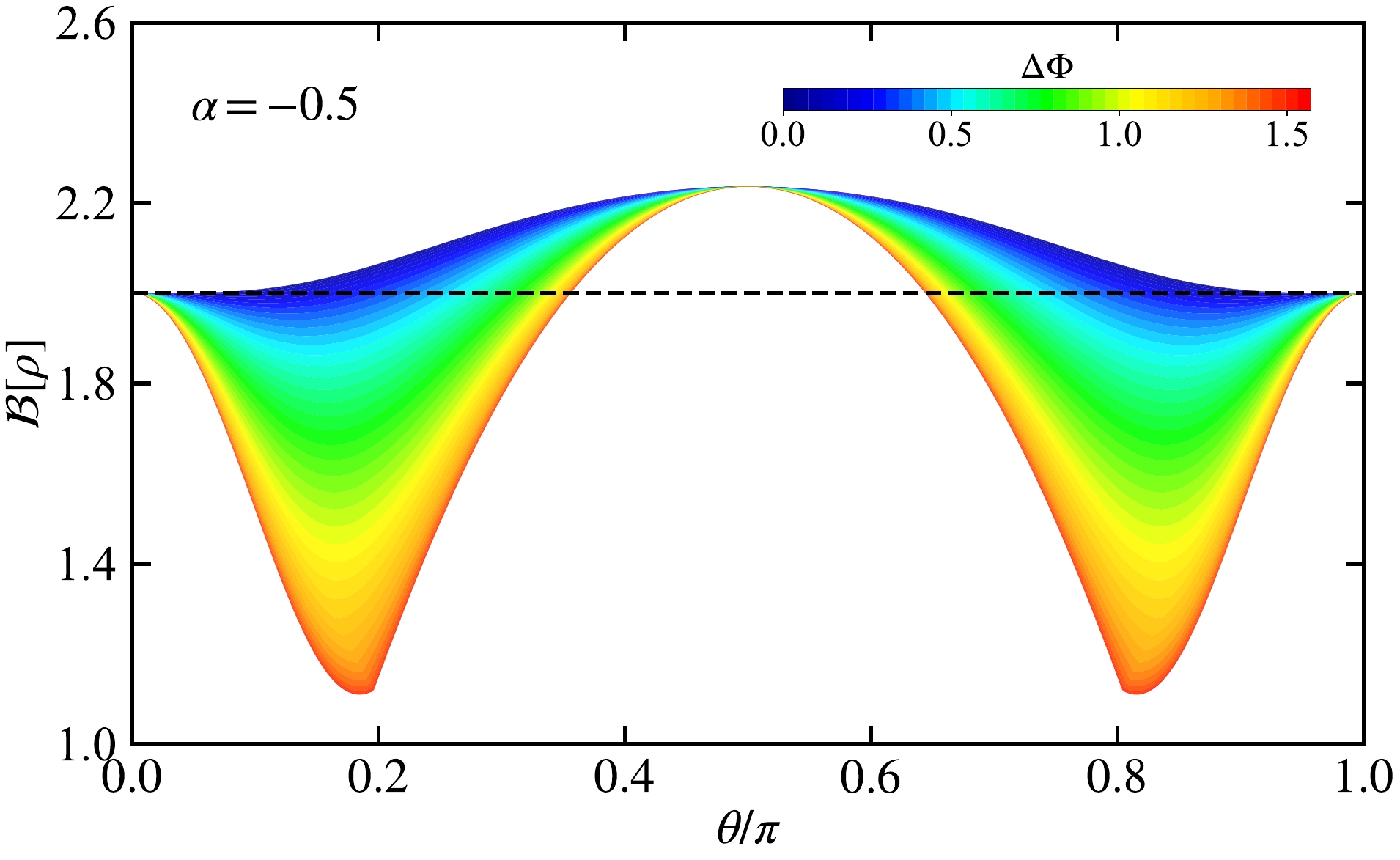}
        \caption{}
    \label{general:d}
    \end{subfigure}
    \hspace{0.027\textwidth}
    \begin{subfigure}{0.3\textwidth}
        \includegraphics[width=\textwidth]{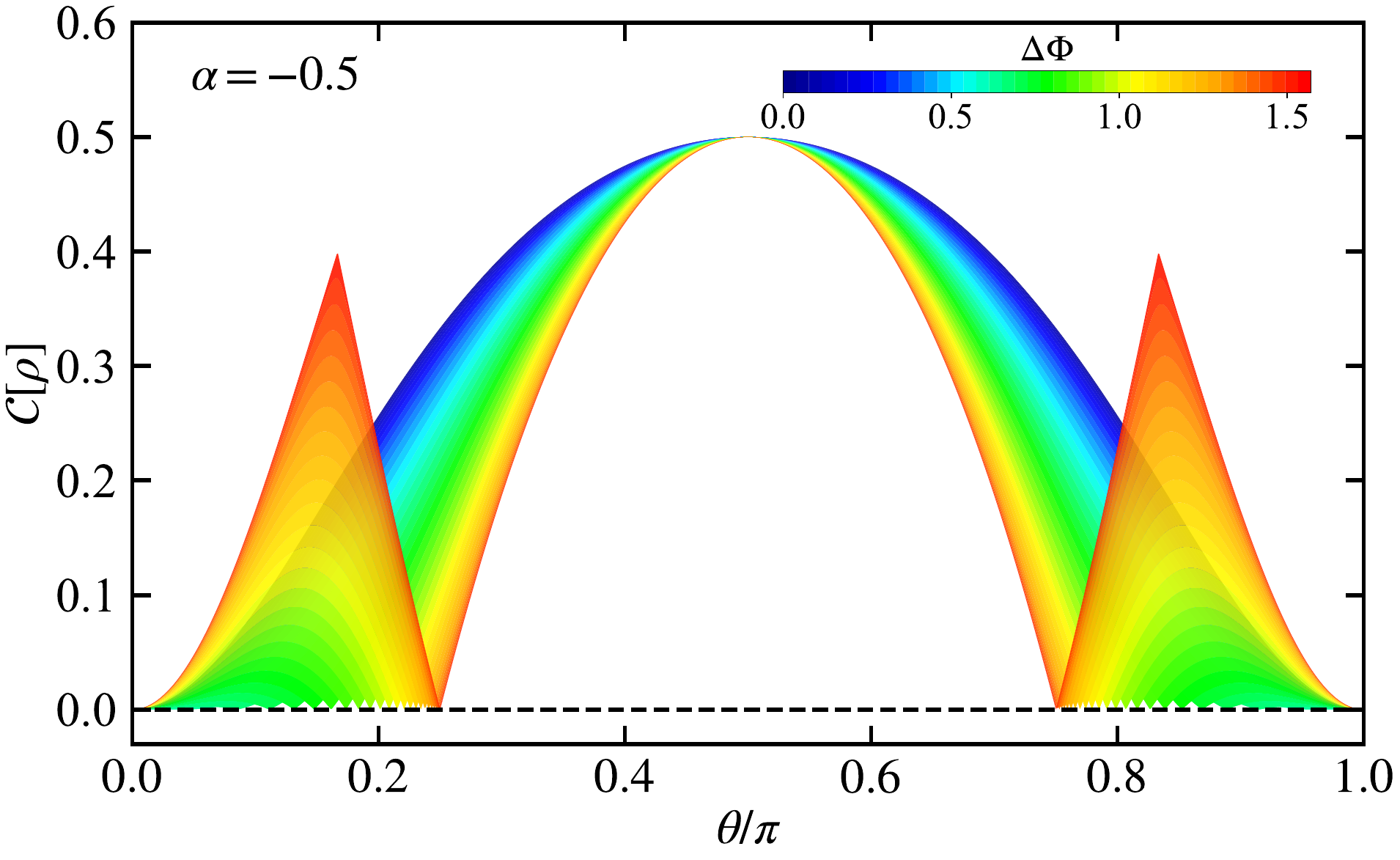}
        \caption{}
    \label{general:e}
    \end{subfigure}
    \hspace{0.027\textwidth}
    \begin{subfigure}{0.3\textwidth}
        \includegraphics[width=\textwidth]{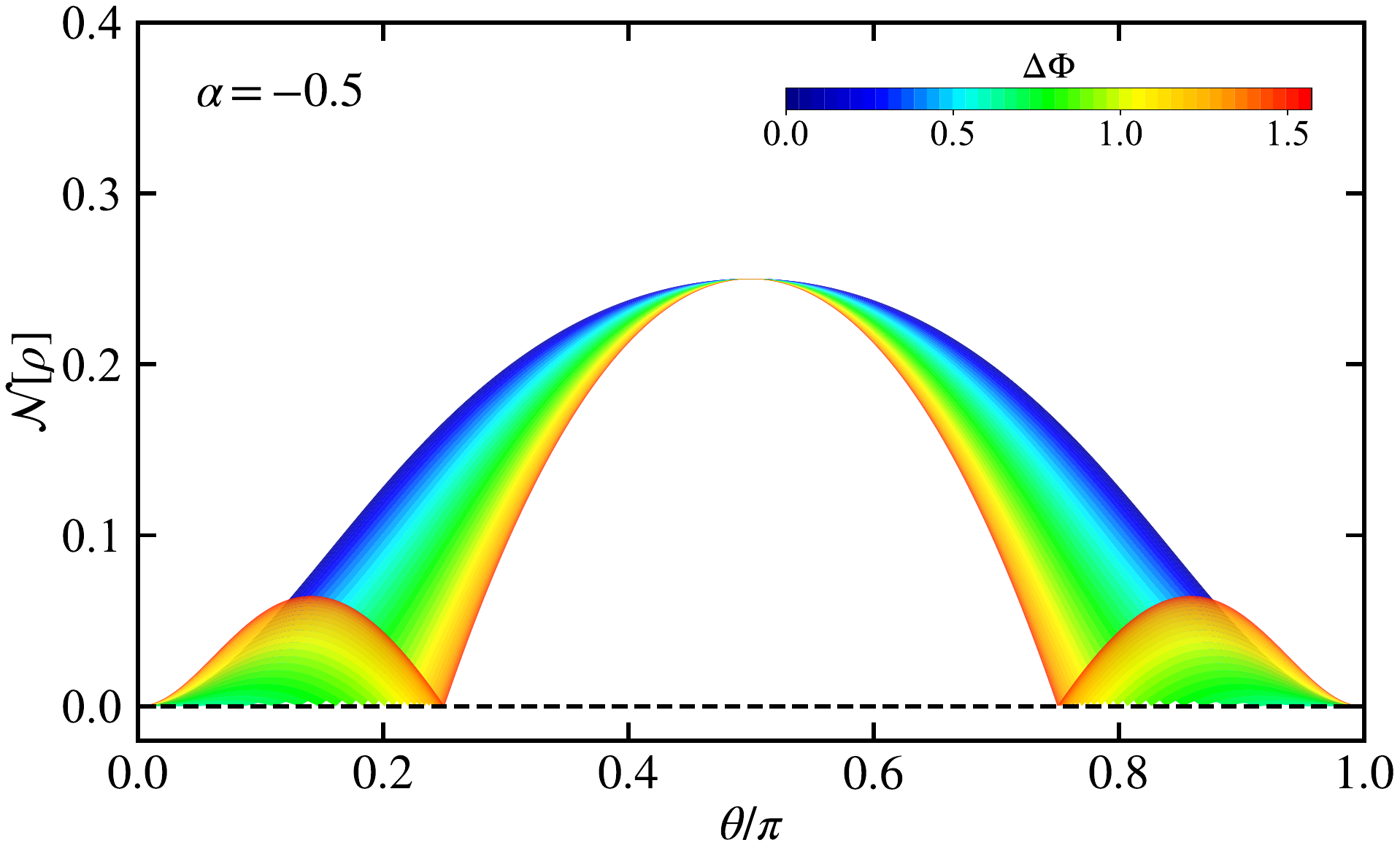}
        \caption{}
    \label{general:f}
    \end{subfigure}
    \vspace{0.2cm}
    \begin{subfigure}[b]{0.3\textwidth}
        \includegraphics[width=\textwidth]{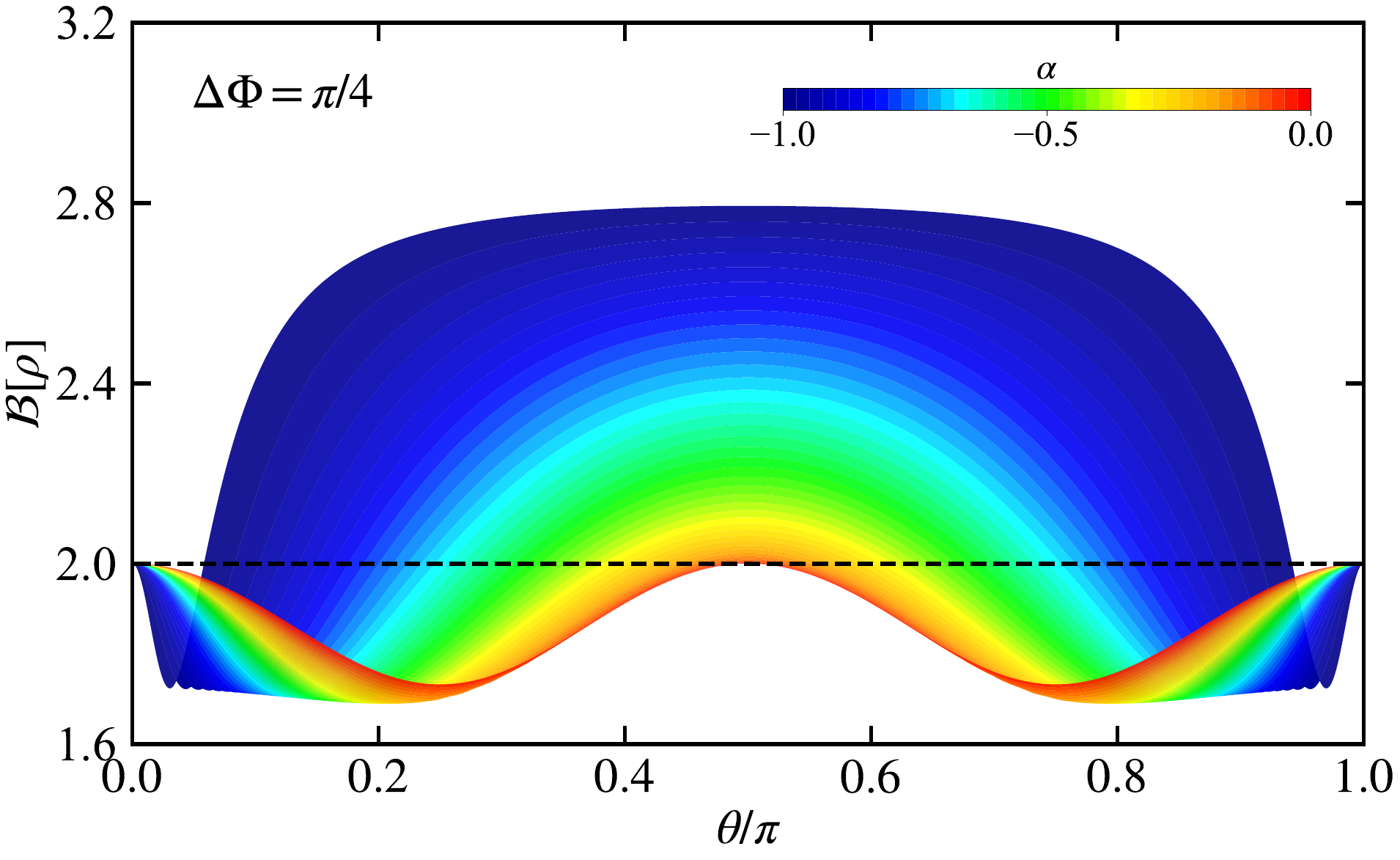}
        \caption{}
    \label{general:g}
    \end{subfigure}
    \hspace{0.027\textwidth}
    \begin{subfigure}[b]{0.3\textwidth}
        \includegraphics[width=\textwidth]{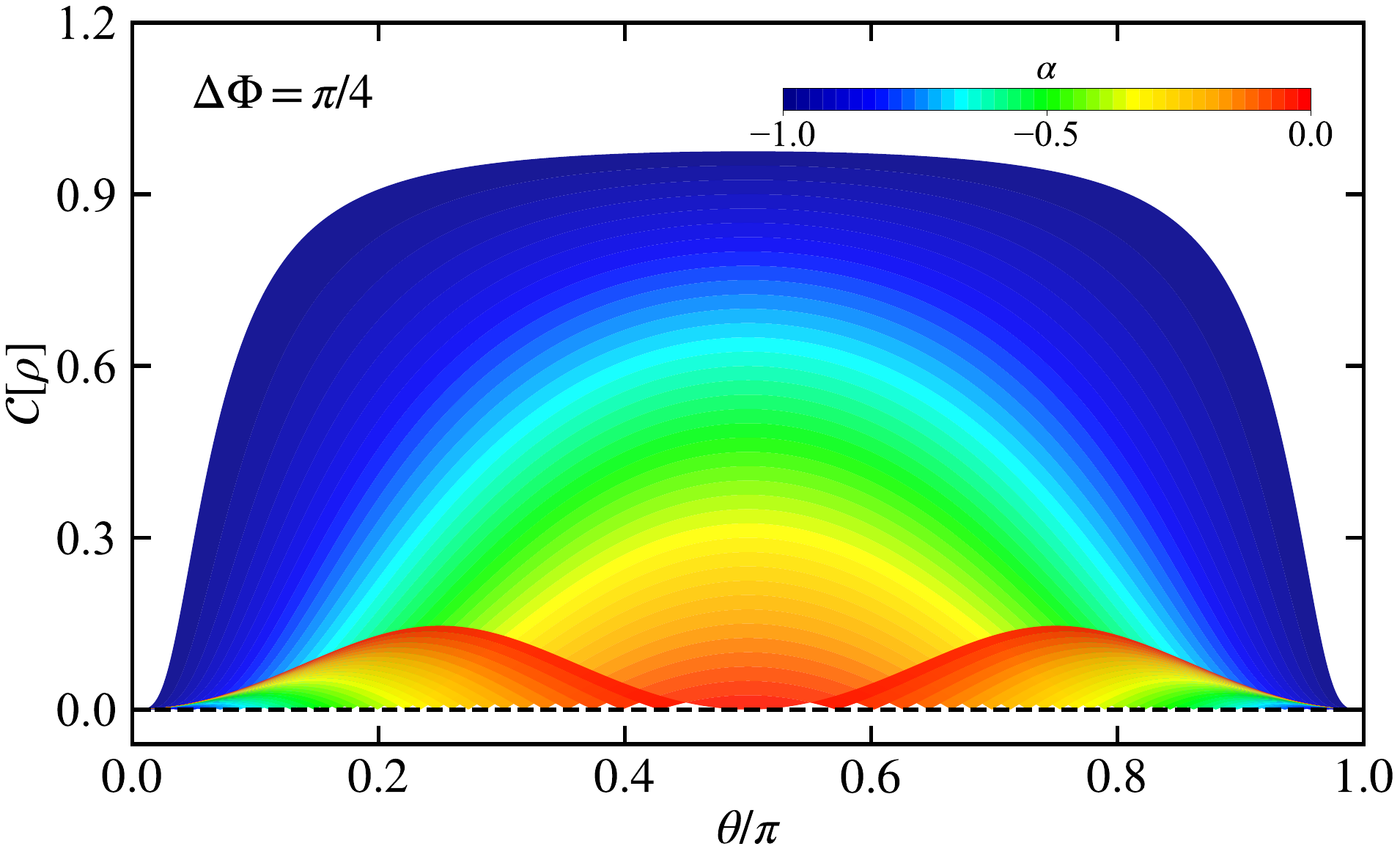}
        \caption{}
    \label{general:h}
    \end{subfigure}
    \hspace{0.027\textwidth}
    \begin{subfigure}[b]{0.3\textwidth}
        \includegraphics[width=\textwidth]{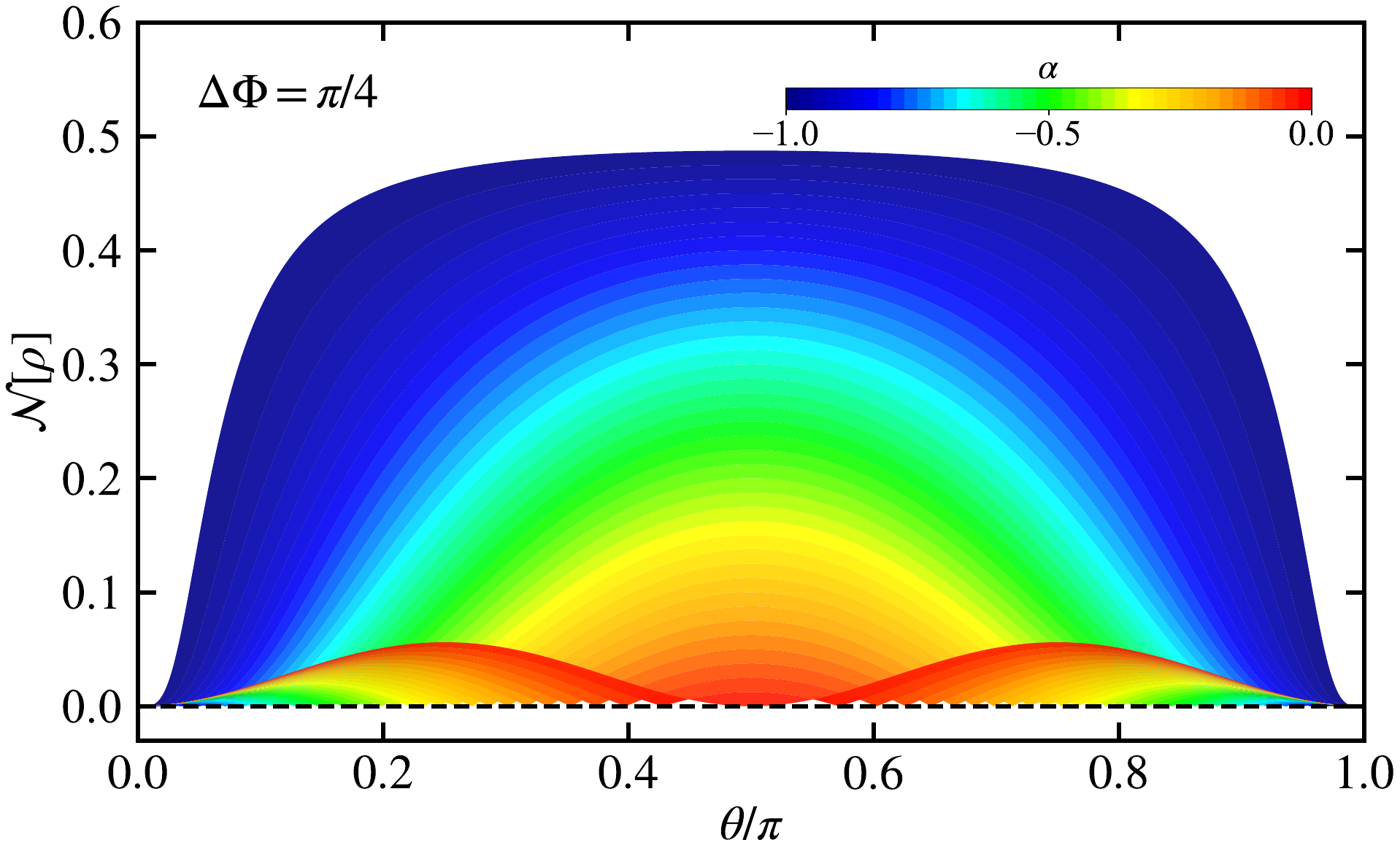}
        \caption{}
    \label{general:i}
    \end{subfigure}
    \vspace{0.2cm}
    \begin{subfigure}[b]{0.3\textwidth}
        \includegraphics[width=\textwidth]{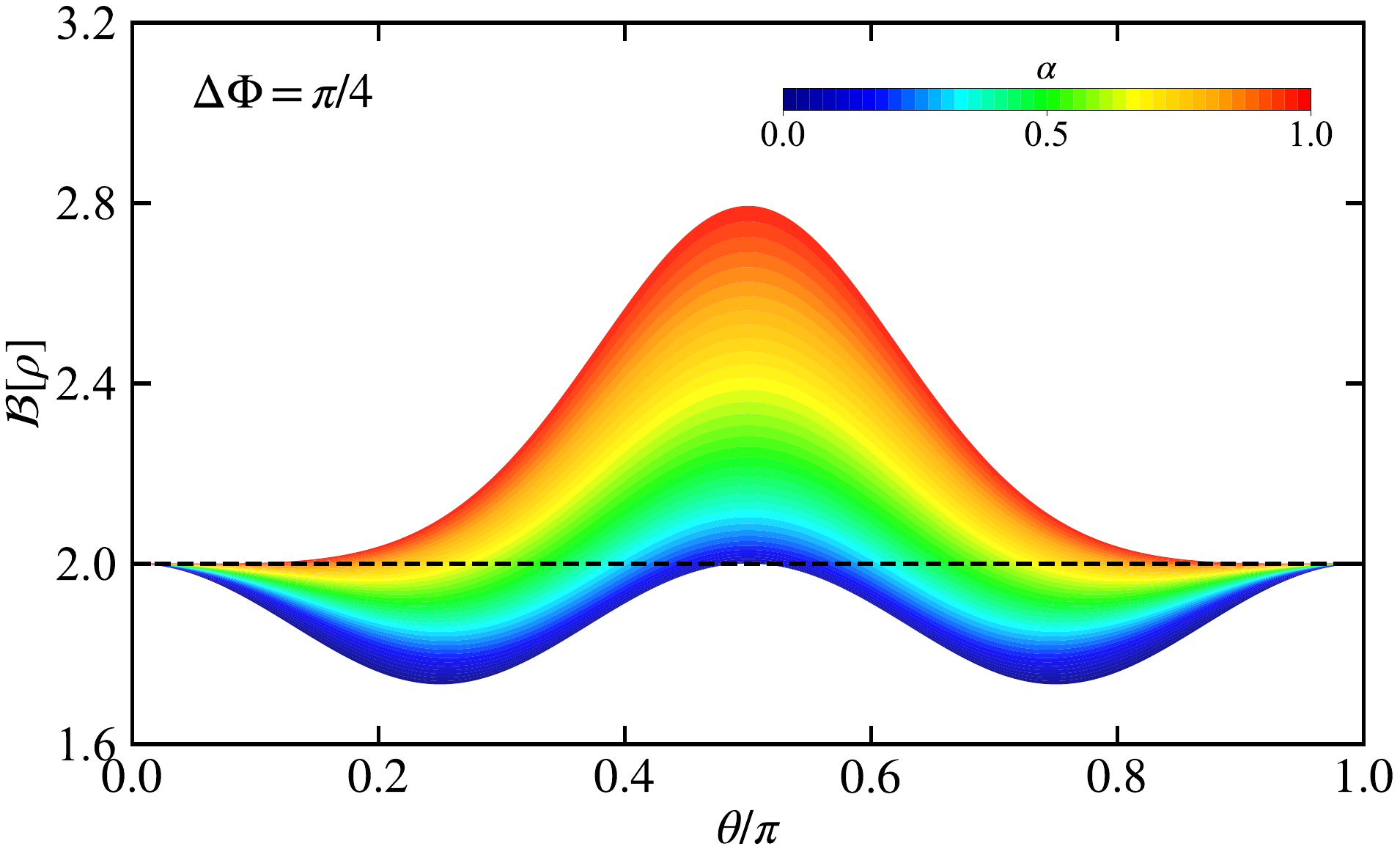}
        \caption{}
    \label{general:j}
    \end{subfigure}
    \hspace{0.027\textwidth}
    \begin{subfigure}[b]{0.3\textwidth}
        \includegraphics[width=\textwidth]{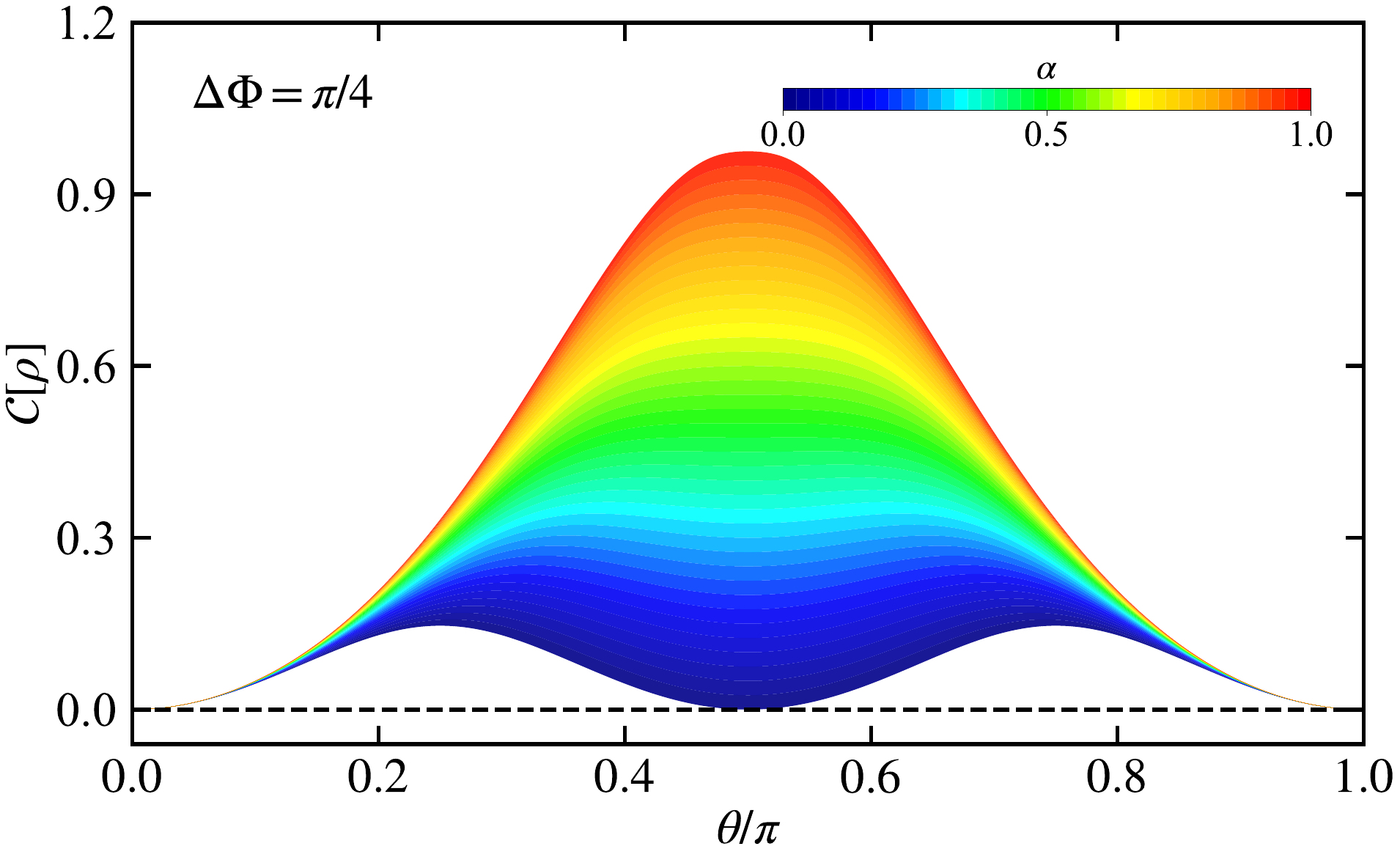}
        \caption{}
    \label{general:k}
    \end{subfigure}
    \hspace{0.027\textwidth}
    \begin{subfigure}[b]{0.3\textwidth}
        \includegraphics[width=\textwidth]{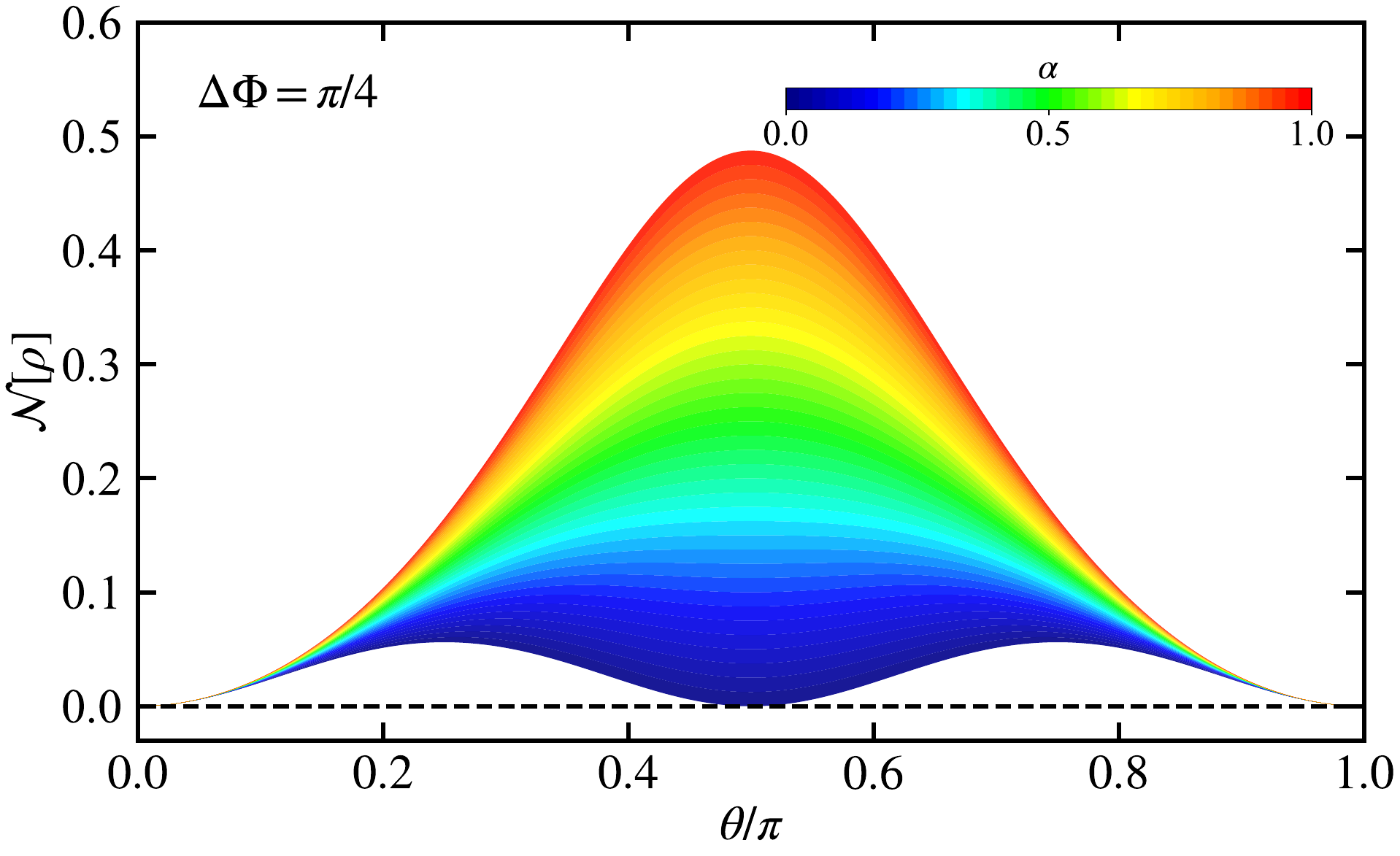}
        \caption{}
    \label{general:l}
    \end{subfigure}
    \vspace{0.2cm}
    \begin{subfigure}[b]{0.3\textwidth}
        \includegraphics[width=\textwidth]{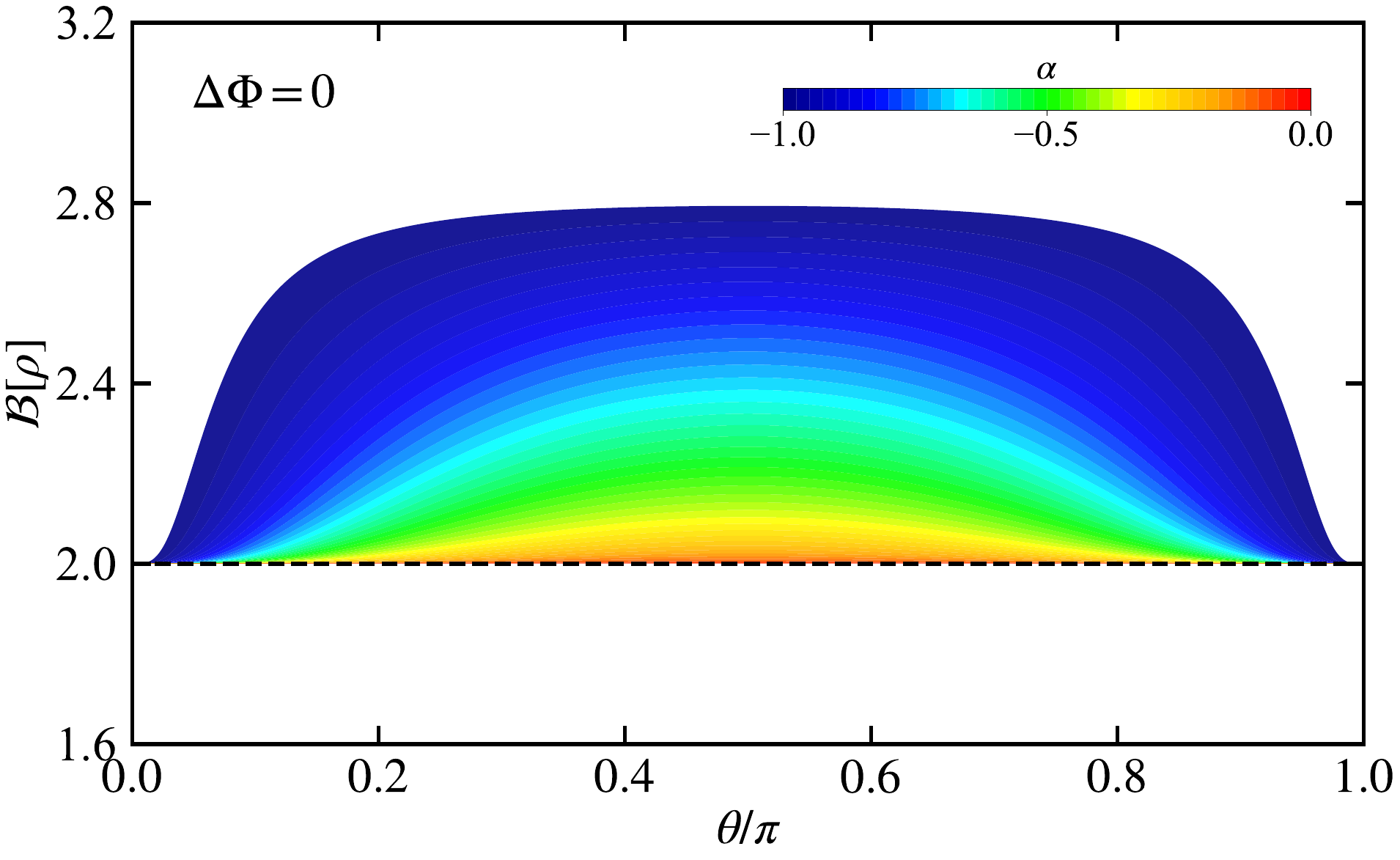}
        \caption{}
    \label{general:m}
    \end{subfigure}
    \hspace{0.027\textwidth}
    \begin{subfigure}[b]{0.3\textwidth}
        \includegraphics[width=\textwidth]{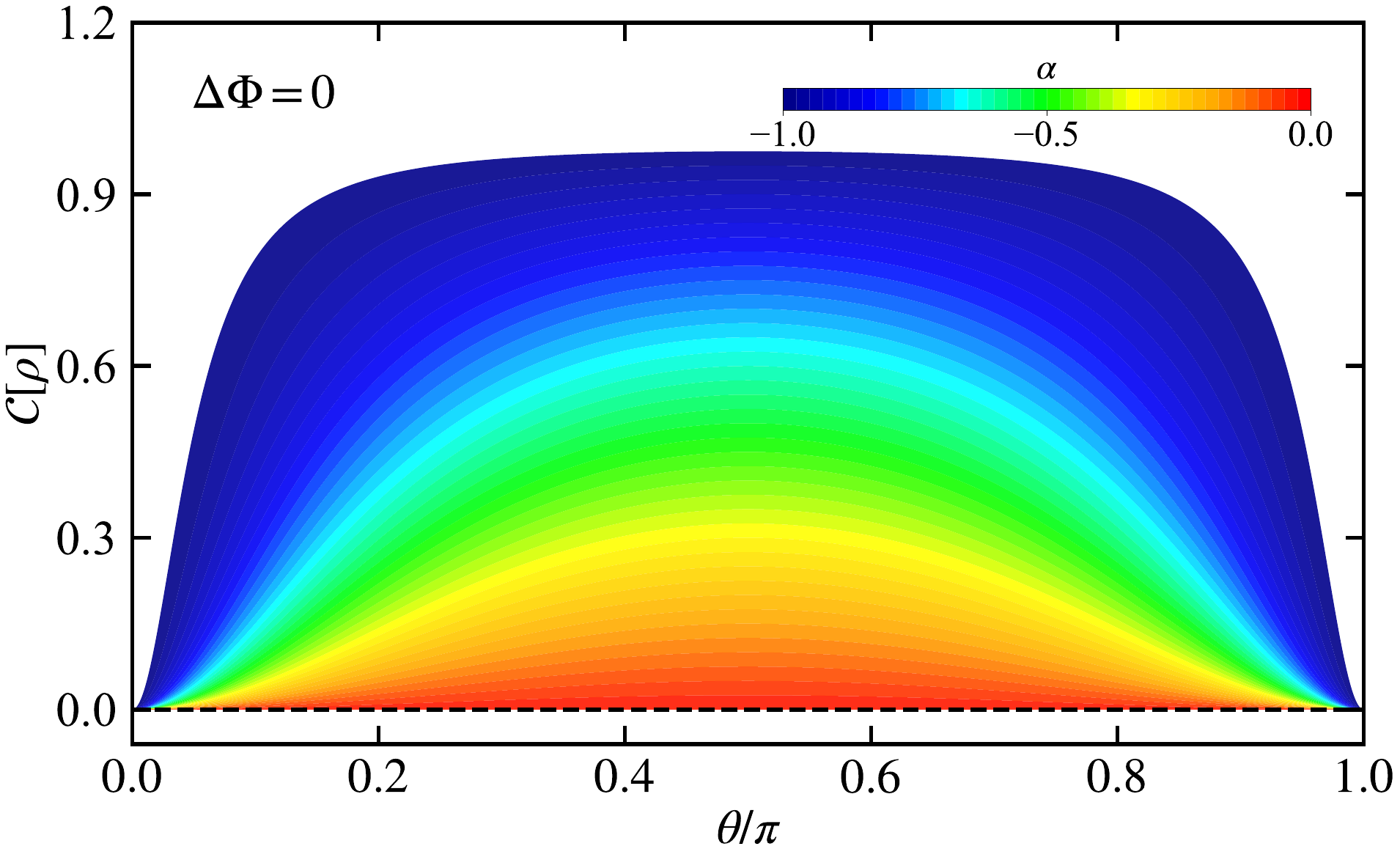}
        \caption{}
    \label{general:n}
    \end{subfigure}
    \hspace{0.027\textwidth}
    \begin{subfigure}[b]{0.3\textwidth}
        \includegraphics[width=\textwidth]{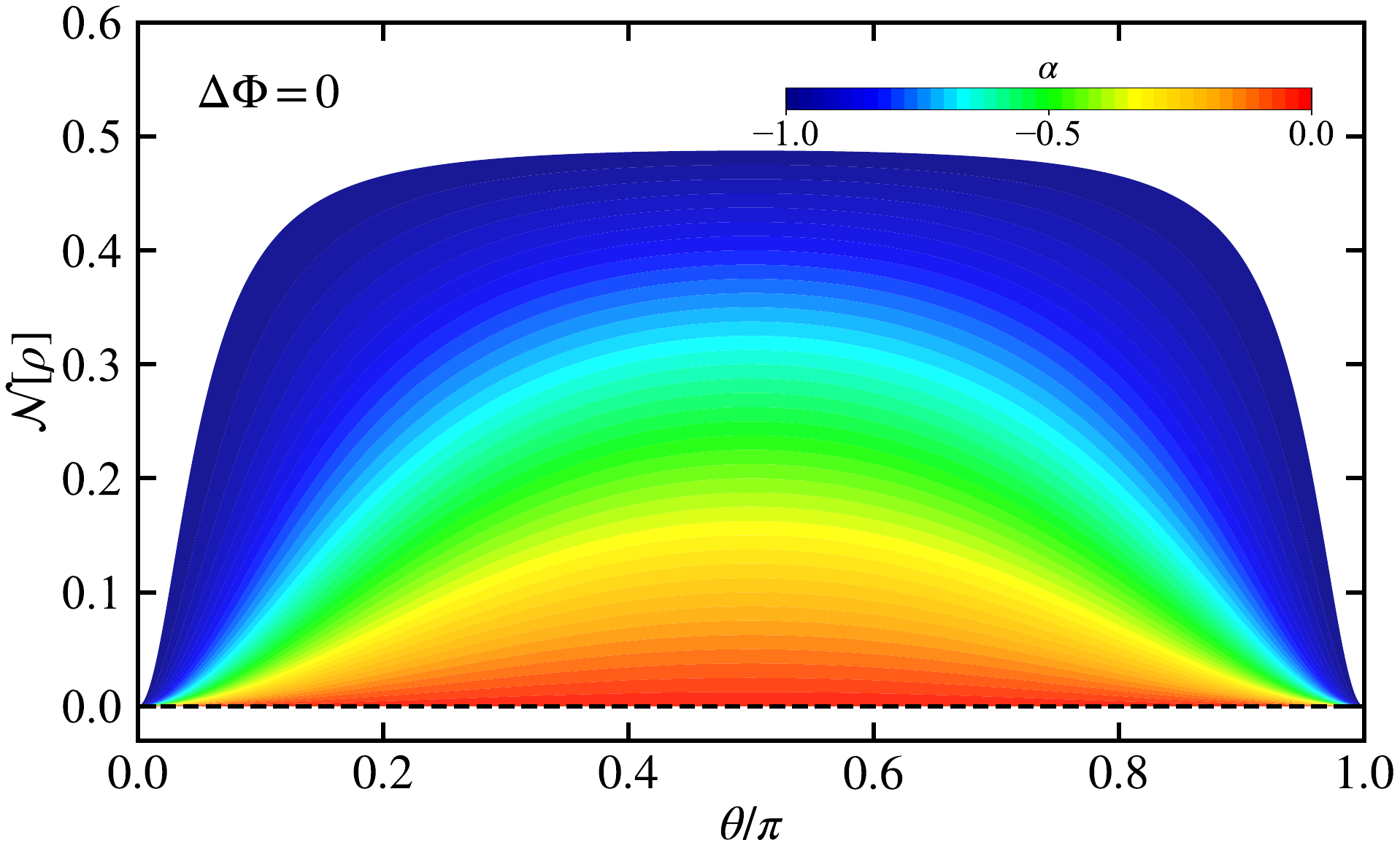}
        \caption{}
    \label{general:o}
    \end{subfigure}
    \vspace{0.2cm}
    \begin{subfigure}[b]{0.3\textwidth}
        \includegraphics[width=\textwidth]{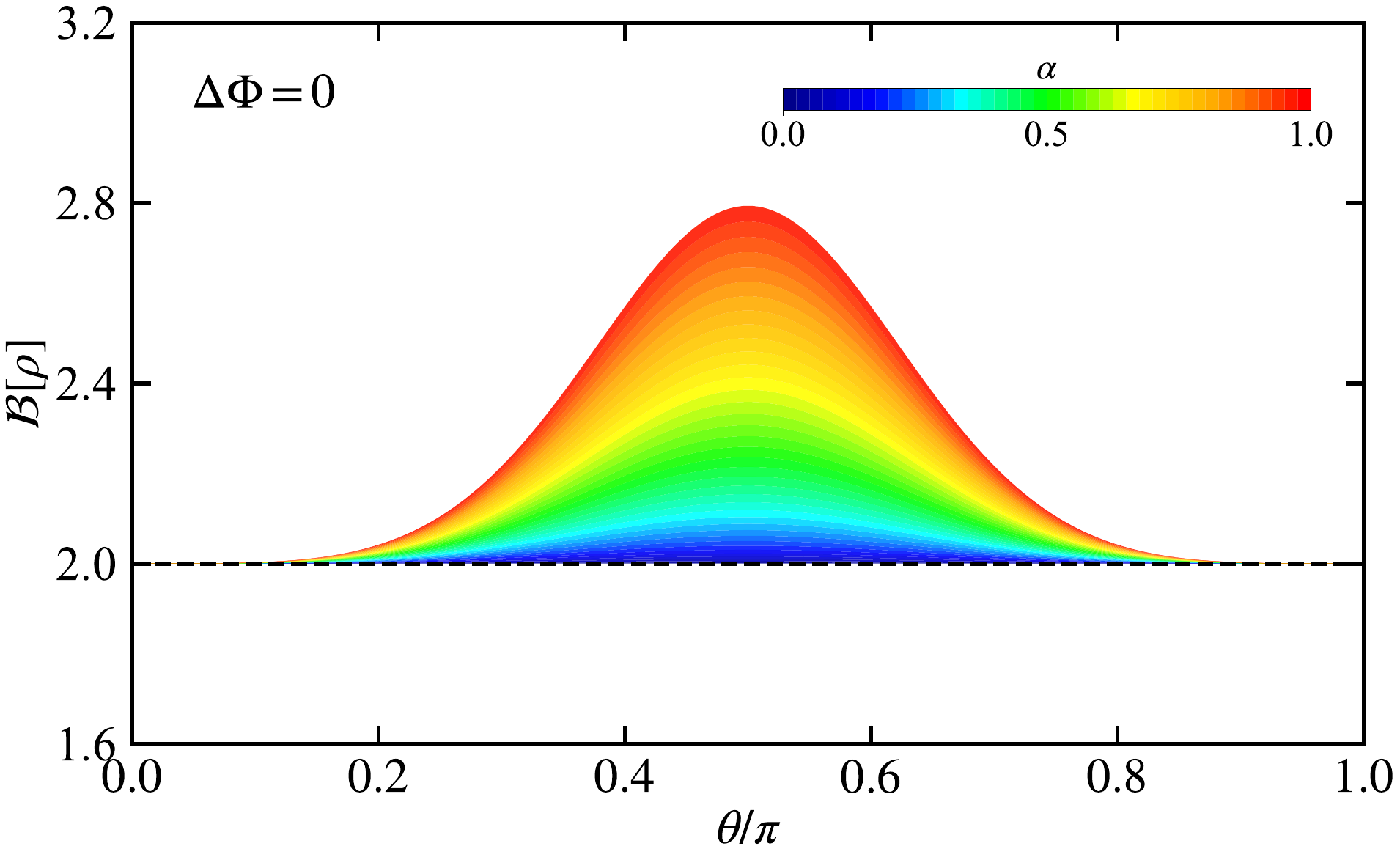}
        \caption{}
    \label{general:p}
    \end{subfigure}
    \hspace{0.027\textwidth}
    \begin{subfigure}[b]{0.3\textwidth}
        \includegraphics[width=\textwidth]{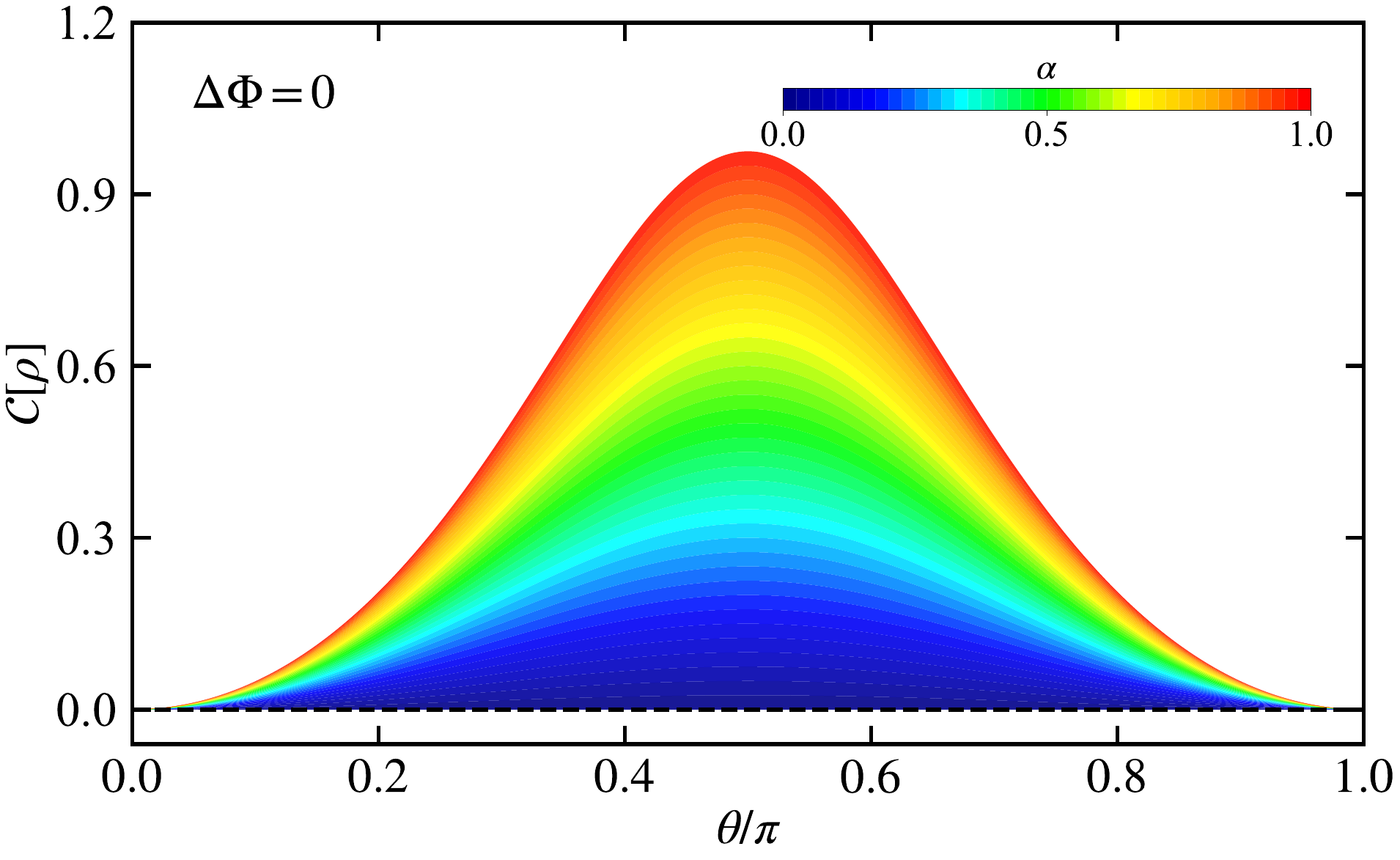}
        \caption{}
    \label{general:q}
    \end{subfigure}
    \hspace{0.027\textwidth}
    \begin{subfigure}[b]{0.3\textwidth}
        \includegraphics[width=\textwidth]{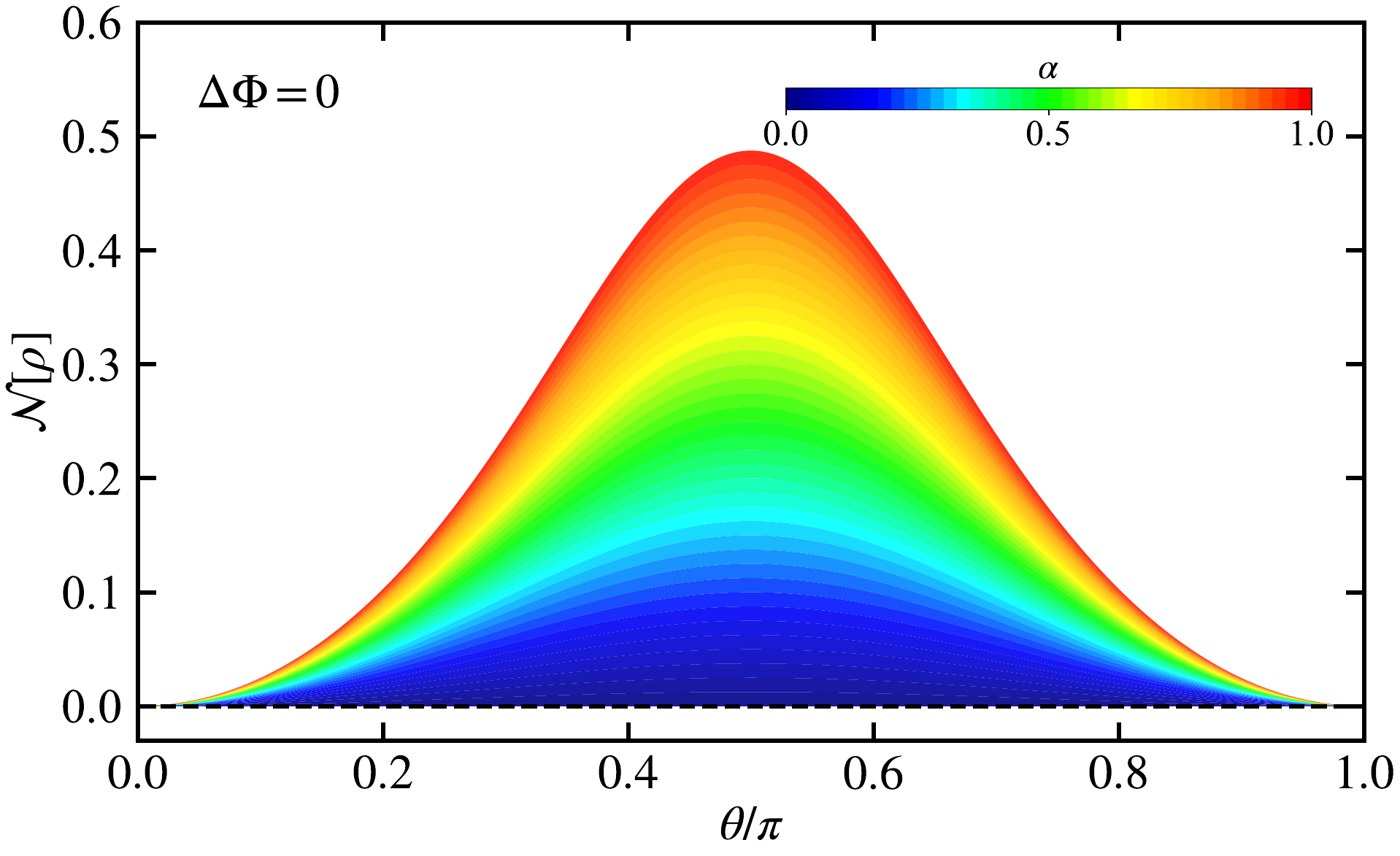}
        \caption{}
    \label{general:r}
    \end{subfigure}
    \caption{Quantum information observables as functions of $\theta$ in different parameters $\alpha$ and $\Delta\Phi$ in $e^+e^-\rightarrow B\bar{B}$ reaction.}
    \label{fig:general-results}
\end{figure}

\subsubsection{Quantum information observables in general systems}

As we can see in Section~\ref{Sec:qtqiob}, for any $B\bar{B}$ system (not limited to the hyperon-antihyperon system) produced in the process $e^+e^-\rightarrow B\bar{B}$, once the parameters $\alpha$ and $\Delta\Phi$ of the reaction are determined, we can directly obtain the spin density operator of $B\bar{B}$ according to Eqs.~(\ref{eqa:spin density operator})-(\ref{eqa:spin-correlation-matrix}), and thereby compute the corresponding quantum information observables. 
For example, $B$ can be a spin-1/2 fundamental particle such as a lepton or a quark, the parameters are $\alpha = 1$ and $\Delta\Phi = 0$, then we can obtain quantum information observables in the same approach. 

Figure~\ref{fig:general-results} displays the quantum information observables in $e^+e^-\rightarrow B\bar{B}$ under various choices of the parameters $\alpha$ and $\Delta\Phi$. In the first column of Fig. \ref{fig:general-results} [(a), (d), (g), (j), (m), (p)] the Bell variables are shown as functions of $\theta$ for different values of $\alpha$ and $\Delta\Phi$, with the black dotted (horizontal) line indicating the threshold value 2. In the second column of Fig. \ref{fig:general-results} [(b), (e), (h), (k), (n), (q)] are presented the results for the concurrence observable $\mathcal{C}$ as functions of $\theta$ under the same parameter sets, where the black dotted (horizontal) line corresponds to the separability bound 0. Similarly, the third column of Fig. \ref{fig:general-results} [(c), (f), (i), (l), (o), (r)] shows the results for the negativity observable $\mathcal{N}$ versus $\theta$, also with the black dotted (horizontal) line marking the bound 0.

According to the first column of Fig.~\ref{fig:general-results}, when $\alpha$ is fixed, the maximum value of $\mathcal{B}$ remains constant which confirms $\mathcal{B}_{\rm max}=\sqrt{1-\alpha^2}$. Moreover, for all values of $\theta$ except $\theta = 0$ and $\theta = \pi/2$, a larger $\Delta\Phi$ in the range $[0, \pi/2]$ leads to smaller $\mathcal{B}$ and smaller range of $\theta$ where the CHSH inequality is violated. When $\Delta\Phi$ is fixed, a larger $|\alpha|$ yields both a generally larger $\mathcal{B}$ and a broader region of $\theta$ where the CHSH inequality is violated. In particular, for $\Delta\Phi = 0$, $\mathcal{B} \geq 2$ always holds or equivalently the CHSH inequality is always violated except at $\theta = 0,\pi$. This demonstrates that the non-zero relative phase of the time-like electric and magnetic form factors, which signals the onset of polarization, is the very reason why the $B\bar{B}$ spins become local at specific range of scattering angle. Consequently, it should be feasible to determine this relative phase experimentally, by observing whether the locality exists in spin states of $B\bar{B}$ for some scattering angles.

Regarding entanglement measures, as shown in the second and third columns of Fig.~\ref{fig:general-results}, when $\alpha\neq 0$, both $\mathcal{C}$ and $\mathcal{N}$ remain positive or in other words the spin states of $B$ and $\bar{B}$ are always entangled at all $\theta$ except some special angles, irrespective of the values of $\alpha$ and $\Delta\Phi$. When $\alpha$ is fixed, both $\mathcal{C}$ and $\mathcal{N}$ have the same value at $\theta=\pi/2$. For $\alpha>0$, increasing $\Delta\Phi$ in $[0,\pi/2]$ enhances both $\mathcal{C}$ and $\mathcal{N}$. For $\alpha<0$, $\mathcal{C}$ and $\mathcal{N}$ increase with increasing $\Delta\Phi$ mainly in the vicinity of $\theta=\pi/2$, while away from this region, they first decrease and then increase with increasing $\Delta\Phi$. When $\Delta\Phi$ is fixed, $\mathcal{C}$ and $\mathcal{N}$ generally increase with $\alpha$.

A comparison of these results reveals that, when the particle has transverse polarization, i.e. $\beta \neq 0$, the behavior of the Bell variable is very different from that of two entanglement measures. However, when the particle is not polarized, the behaviors of these quantum information observables are similar. Furthermore, the concurrence and negativity show a high degree of qualitative similarity in all cases considered, suggesting that the negativity can serve as a trustworthy measure of the entanglement in higher spin particle systems, such as spin-1 or spin-3/2 particle-antiparticle systems.

\subsubsection{Implication for parton entanglement in quarkonium} 

The methodology presented in this paper also provides theoretical insight for the experimental measurement of spin entanglement between valence quarks inside heavy-flavored mesons. 

As recently investigated in Ref.~\cite{Zhang:2026pff} using a phenomenological model within the light-cone Hamiltonian approach~\cite{Vary:2025yqo}, one can explore the quantum entanglement between a quark and antiquark in heavy quarkonium systems. In the reaction $e^+e^-\rightarrow\Lambda^+_c\bar{\Lambda}^-_c$, an intermediate vector meson state $\psi^*$ composed of $c\bar{c}$ is produced at appropriate collision energies. According to the OZI rule, the decay process $\psi^*\rightarrow\Lambda^+_c\bar{\Lambda}^-_c$ is predominantly governed by the ``heavy quark line'' mechanism. In this picture, the $c$ and $\bar{c}$ quarks from the initial charmonium state barely participate in any additional interactions and instead transition directly into the final-state hyperons.
Crucially, in the heavy-quark limit, the two light quarks ($u$ and $d$) inside the $\Lambda^+_c$ form a spin-0 (scalar) diquark, meaning the spin of the $\Lambda^+_c$ is almost entirely determined by the $c$ quark. During the non-perturbative hadronization process, Heavy Quark Spin Symmetry (HQSS) dictates that the spin degree of freedom of the heavy quark decouples from the light-quark dynamics. Because this decoupling suppresses heavy-quark spin-flip transitions by a factor of $\Lambda_{\text{QCD}}/m_c$, the $c$ and $\bar{c}$ quarks largely retain the spin information they originally held within the charmonium state.

Therefore, one can infer that the observable spin correlation between the $\Lambda^+_c$ and $\bar{\Lambda}^-_c$ in the process $e^+e^-\rightarrow\psi^*\rightarrow\Lambda^+_c\bar{\Lambda}^-_c$ essentially reflects the primordial spin entanglement between the $c$ and $\bar{c}$ quarks inside the $\psi^*$. This preservation mechanism provides a feasible and robust experimental pathway for probing parton-level spin entanglement within heavy-flavored mesons.

\section{Summary and outlook}
\label{Sec:Summary and outlook}

We investigate several quantum information observables including the Bell variable, negativity and concurrence in $B\bar{B}$ systems produced in $e^+e^-$ annihilation, where $B$ denotes the spin-1/2 particle. Using the parameters $\alpha$ and $\Delta\Phi$ measured at BESIII experiments, we present the numerical results for these observables in hyperon-antihyperon systems. We explore the relation between the Bell variable and entanglement measures, as well as the relation between two different entanglement measures concurrence and negativity. We also study the behaviors of these observables in general cases. 

Firstly, we observe that there is no strong dependence of these quantum information observables on $\sqrt{s}$ at BESIII experiments. Secondly, we find that in characterizing the spin correlation between $B$ and $\bar{B}$, the Bell variable is very different from entanglement measures such as concurrence and negativity. The Bell variable serves specifically as a measure for nonlocality and is not suitable for quantifying the magnitude of the entanglement. In contrast, concurrence and negativity has similarity in quantifying the quantum entanglement and can be regarded as effective measures for the entanglement. Finally, we also observe that the polarization of $B$ and $\bar{B}$ not only induces the locality but also influences the behavior of the entanglement measures.

This work, together with other similar studies~\cite{Wu:2024asu,Wu:2025dds,Zhang:2026nwm,Li:2026bkf}, provides a comprehensive picture for quantum correlation in spin-1/2 particle-antiparticle systems produced in $e^+e^-$ annihilation. The application of quantum information tools, such as the Bell variable, concurrence and negativity, to high-energy physics processes opens a new possibility for the experimental investigation of hadron structure. 

{\bf Acknowledgement}. The authors thank H.B. Li for suggesting us to study $\Lambda^+_c\bar{\Lambda}^-_c$ system and for helpful discussion. This work is supported in part by the Chinese Academy of Sciences under Grant No. YSBR-101. Q.W. is supported by the National Natural Science Foundation of China under Grant No.~12135011. 

\appendix
\section{Derivation of the analytical formulas of quantum information observables}
\label{appendix:formula-ob}
To analytically obtain the quantum information observables, it is convenient to transform the two-qubit density operator (\ref{eqa:spin density operator}) into the $X$-state by using a local unitary transformation~\cite{Wu:2024asu}. After the transformation, the operator becomes
\begin{equation}
    \rho_{B\bar{B}}^X=\frac{1}{4}\Big(\mathbbmtt{1}\otimes\mathbbmtt{1}+a\sigma_z\otimes\mathbbmtt{1}
    +a\mathbbmtt{1}\otimes\sigma_z+\sum_it_i\sigma_i\otimes\sigma_i\Big),
\label{eqa:X-state}
\end{equation}
where
\begin{equation}
\begin{split}
    &a=\frac{\beta\operatorname{sin}\theta\operatorname{cos}\theta}{1+\alpha\operatorname{cos}^2\theta},\\
    t&_{1,2}=\frac{1+\alpha\pm\sqrt{(1+\alpha\operatorname{cos}2\theta)^2-\beta^2\operatorname{sin}^22\theta}}{2(1+\alpha\operatorname{cos}^2\theta)},\\
    &t_3=-\frac{\alpha\operatorname{sin}^2\theta}{1+\alpha\operatorname{cos}^2\theta}.
\label{eqa:X-state elements}
\end{split}
\end{equation}
Since the quantum correlation in a bipartite system keeps invariant under the local unitary transformation~\cite{Dur:2000zz}, the density operator $\rho_{B\bar{B}}$ and $\rho_{B\bar{B}}^X$ derive the identical quantum information observables.
By explicitly express the Pauli matrices in (\ref{eqa:X-state}), the $X$-state density operator can be written into the form
\begin{equation}
\begin{split}
    \rho_{B\bar{B}}^{X}=\frac{1}{4}
    \begin{pmatrix}
        1+2a+t_3 & 0 & 0 & t_1-t_2 \\
        0 & 1-t_3 & t_1+t_2 & 0 \\
        0 & t_1+t_2 & 1-t_3 & 0 \\
        t_1-t_2 & 0 & 0 & 1-2a+t_3
    \end{pmatrix}.
\label{eqa:expansion of the X-state}
\end{split}
\end{equation}
Since we have put the density operator into the $X$ form, the correlation matrix is diagonal: $\operatorname{diag}\{t_1,t_2,t_3\}$. The three eigenvalues of $C^TC$ are just $t_1^2,t_2^2$ and $t_3^2$. Since $t_1^2\geq t_2^2$ always satisfies, according to (\ref{eqa:Bell variable}), the Bell variable is
\begin{equation}
    \mathcal{B}[\rho_{B\bar{B}}^{X}]=2\sqrt{\operatorname{max}\{t_1^2+t_2^2,t_1^2+t_3^2\}},
\end{equation}
where $t_1^2+t_2^2$ and $t_1^2+t_3^2$ are given by
\begin{equation}
    t_1^2+t_2^2=1+\bigg(\frac{\alpha\operatorname{sin}^2\theta}{1+\alpha\operatorname{cos}^2\theta}\bigg)^2-2\bigg(\frac{\beta\operatorname{sin}\theta\operatorname{cos}\theta}{1+\alpha\operatorname{cos}^2\theta}\bigg)^2,
\end{equation}
\begin{equation}
    t_1^2+t_3^2=\Bigg(\frac{1+\alpha+\sqrt{(1+\alpha\operatorname{cos}2\theta)^2-\beta^2\operatorname{sin}^22\theta}}{2(1+\alpha\operatorname{cos}^2\theta)}\Bigg)^2+\frac{\alpha^2(1-\operatorname{cos}2\theta)^2}{4(1+\alpha\operatorname{cos}^2\theta)^2}.
\end{equation}
According to Ref.~\cite{Yu:2007bxs}, the concurrence for the $X$-state is given by
\begin{equation}
    \mathcal{C}[\rho^X]=2\operatorname{max}\Big\{0,|\rho_{14}^X|-\sqrt{\rho_{22}^X\rho_{33}^X},|\rho_{23}^X|-\sqrt{\rho_{11}^X\rho_{44}^X}\Big\}.
\end{equation}
With (\ref{eqa:X-state elements}) and (\ref{eqa:expansion of the X-state}), the analytical formula of the concurrence for the $B\bar{B}$ system is
\begin{equation}
    \mathcal{C}[\rho_{B\bar{B}}^X]=|t_2|=\frac{\big|1+\alpha-\sqrt{(1+\alpha\operatorname{cos}2\theta)^2-\beta^2\operatorname{sin}^22\theta}\big|}{2(1+\alpha \operatorname{cos}^2\theta)}
\end{equation}
By partially transpose the bases of $\bar{B}$ in density operator (\ref{eqa:X-state}), we can obtain the partially transposed density operator of the $B\bar{B}$ system
\begin{equation}
\begin{split}
    (\rho_{B\bar{B}}^X)^{T_{\bar{B}}}&=\frac{1}{4}\Big(\mathbbmtt{1}\otimes\mathbbmtt{1}^T+a\sigma_z\otimes\mathbbmtt{1}^T
    +a\mathbbmtt{1}\otimes\sigma_z^T+\sum_it_i\sigma_i\otimes\sigma_i^T\Big)\\
    &=\frac{1}{4}
    \begin{pmatrix}
        1+2a+t_3 & 0 & 0 & t_1+t_2 \\
        0 & 1-t_3 & t_1-t_2 & 0 \\
        0 & t_1-t_2 & 1-t_3 & 0 \\
        t_1+t_2 & 0 & 0 & 1-2a+t_3
    \end{pmatrix}.
\end{split}
\end{equation}
The negativity is
\begin{equation}
    \mathcal{N}(\rho_{B\bar{B}}^{X})=\frac{1}{2}\bigg(\sum_{i=1}^4|\lambda_i|-1\bigg),
\end{equation}
where
\begin{equation}
\begin{split}
    \lambda_{1,2}=&\frac{1}{4}\bigg(1+t_3\pm\sqrt{4a^2+(t_1+t_2)^2}\bigg)\\
    &\lambda_3=\frac{1}{4}(1+t_1-t_2-t_3)\\
    &\lambda_4=\frac{1}{4}(1-t_1+t_2-t_3)
\end{split}
\end{equation}
are the eigenvalues of the partially transposed density operator.

\bibliographystyle{unsrt}
\bibliography{ref}

\end{document}